\documentclass[
4pt,
superscriptaddress,
preprint,
 amsmath,amssymb,
 aps,
pra,
]{revtex4-1}

\usepackage{graphicx}
\usepackage{dcolumn}
\usepackage[colorlinks=true,
            linkcolor=blue,
            urlcolor=black,
            citecolor=blue]{hyperref}
\usepackage{bm}
\usepackage{enumerate}
\usepackage{lipsum}
\usepackage{threeparttable}
\usepackage{epstopdf}
\usepackage{float}
\usepackage{xcolor,colortbl}
\usepackage[caption = false]{subfig}
\usepackage{tabularx}

\begin{document}


\title{Tunable  magic wavelengths for trapping with focused Laguerre-Gaussian beams}
\author{Anal Bhowmik}
\email{analbhowmik@phy.iitkgp.ernet.in}
\affiliation{Department of Physics, Indian Institute of Technology Kharagpur, Kharagpur-721302, India.}
\author{Narendra Nath Dutta}
\email{nnd0004@auburn.edu}
\affiliation{Department of Chemistry and Biochemistry, Auburn University, AL-36849, USA.}
\author{Sonjoy Majumder}
\email{sonjoym@phy.iitkgp.ernet.in}
\affiliation{Department of Physics, Indian Institute of Technology Kharagpur, Kharagpur-721302, India.}

\date{\today}





\begin{abstract}

We  present in this paper a theory of dynamic polarizability for an atomic state due to an external field of non-paraxial Laguerre-Gaussian (LG) beam using the sum-over-states technique. A highly correlated relativistic coupled-cluster theory is used to evaluate the most important and correlation sensitive parts of the sum.  The theory is applied on Sr$^+$ to determine the magic wavelengths for  $5s_{{1}/{2}}\rightarrow 4d_{{3}/{2},  {5}/{2}}$ transitions. Results show the variation of magic wavelengths with the choice of orbital and spin angular momenta   of the incident  LG beam. Also, the tunability of the magic wavelengths is studied using the focusing angle of the LG beam and observed its efficiency in the near-infrared region. Evaluations of the wide spectrum  of magic wavelengths from infrared to ultra-violet  have substantial importance to the experimentalists for carrying out high precision measurements in fundamental physics. These  magic wavelengths can be used to confine the atom or ion at  the dark central node or at the high-intensity  ring  of  the  LG beam.

\end{abstract}

\maketitle
\section{INTRODUCTION}
Mechanisms of cooling and trapping of atoms or ions using laser beam have been widely employed  in high precision spectroscopic measurements. To minimize the various systematics in the measurements of any spectroscopic properties \cite{Champenois2004, Chou2010},  experimentalists  need to trap the atoms at particular wavelengths of the external laser field where the differential ac stark shift of an atomic transition effectively vanishes. These  special wavelengths are named as magic wavelengths and are used as to perform clock frequency  measurements \cite{Margolis2009}, optical frequency standards \cite{Ovsiannikov2005}, etc.  Another significant  application of magic wavelengths is in quantum computation and communication schemes when the neutral atoms are trapped inside high-Q cavities at magic wavelengths in the strong-coupling regime \cite{McKeever2003}.

Alkali atoms are favorable candidates for performing experiments using laser cooling and trapping techniques. This is mainly as the low-lying transitions for these atoms are easily accessible by the available laser sources. Stellmer \textit{et al.} \cite{Stellmer2009} produced Bose-Einstein Condensation of $^{88}$Sr using evaporative cooling and optical dipole trap.  A detailed study of sub-Doppler cooling of  $^{87}$Sr in a magneto-optic trap (MOT) has already been investigated \cite{Xu2003}.  This fermionic isotope   is one of the highest-quality, neutral-atom-based optical frequency standards with accuracy below $10^{-18}$ second \cite{Nicholson2015}. The 	distinctive property of this atomic clock is that the atoms are trapped at the  magic wavelengths of an external laser field.

Determination of the magic wavelengths of alkali-metal atoms for linearly polarized (i.e., spin angular momentum (SAM) equal to zero) laser sources has been well explored in literature \cite{Arora2007, Ludlow2008}. Compared to the linearly polarized light,  circularly polarized light (i.e., SAM=$\pm 1$) has an extra part of the total polarizability, called the vector part which arises due to  the dipole moment perpendicular to the  field. For the circularly polarized light, this vector part   has some advantages in the evaluation of the valence polarizability \cite{Flambaum2008, Arora2012}.

 Most of the previous works in the area of trapping have assumed plane-wave or Gaussian modes of a laser. Kuga \textit{et al.} \cite{Kuga1997} were first realized Laguerre-Gaussian (LG) based dipole trap and were confined 10$^8$ numbers of rubidium atoms to the core of a blue-detuned vortex beam (see FIG. 1). Several recent experimental explorations  of trapping atoms using  LG light beams \cite{Otsu2014, Kiselev2016,  Kennedy2014} suggest the importance of the process. The distinct spatial intensity profile  of this LG beam  carries a phase singularity on  its axis \cite{Allen1992, Mondal2014}. This beam is associated with  orbital angular momentum (OAM) due to the helical phase front \cite{Mondal2014, Bhowmik2016}. As shown in FIG. 1, trapping of an atom is possible in the bright or dark region of an LG beam depending upon the sign of detuning. Apart from OAM, the LG beam also carries SAM associated with its  polarization.  During the interaction of paraxial LG beam with atoms or ions, which is  below its recoil limit,  the lowest order transition is possible at quadrupole level \cite{Mondal2014,  Schmiegelow2016} where the electronic motion is affected by the OAM of the LG beam. Therefore,  the OAM of paraxial LG beam does not influence dipole polarizability of an atomic state, but certainly has some effect on  quadrupolar polarizability. Hence the dipole polarizability solely depends on the SAM of the paraxial LG beam.  But the situation is different when the non-paraxial or focused circular LG beam is considered. Here, the SAM and OAM of the beam are coupled,  and they are not conserved separately  \cite{Monteiro2009,Bhowmik2016}. But the total angular momentum (=OAM+SAM) is conserved in interaction with atom \cite{Marrucci2006,Zhao2007}. In our recent work \cite{Bhowmik2016}, we have shown that along with the SAM,  the OAM of a focused LG beam can be transferred to the electronic motion  of cold atoms in the dipole approximation level. This leads to  OAM- and SAM-dependent dipole polarizability of an atomic state and magic wavelengths of a transition.    \\        
 
 \begin{figure*}[!h]
{\includegraphics[trim = 0.5cm 0.5cm 8.1cm 18.5cm, scale=.80]{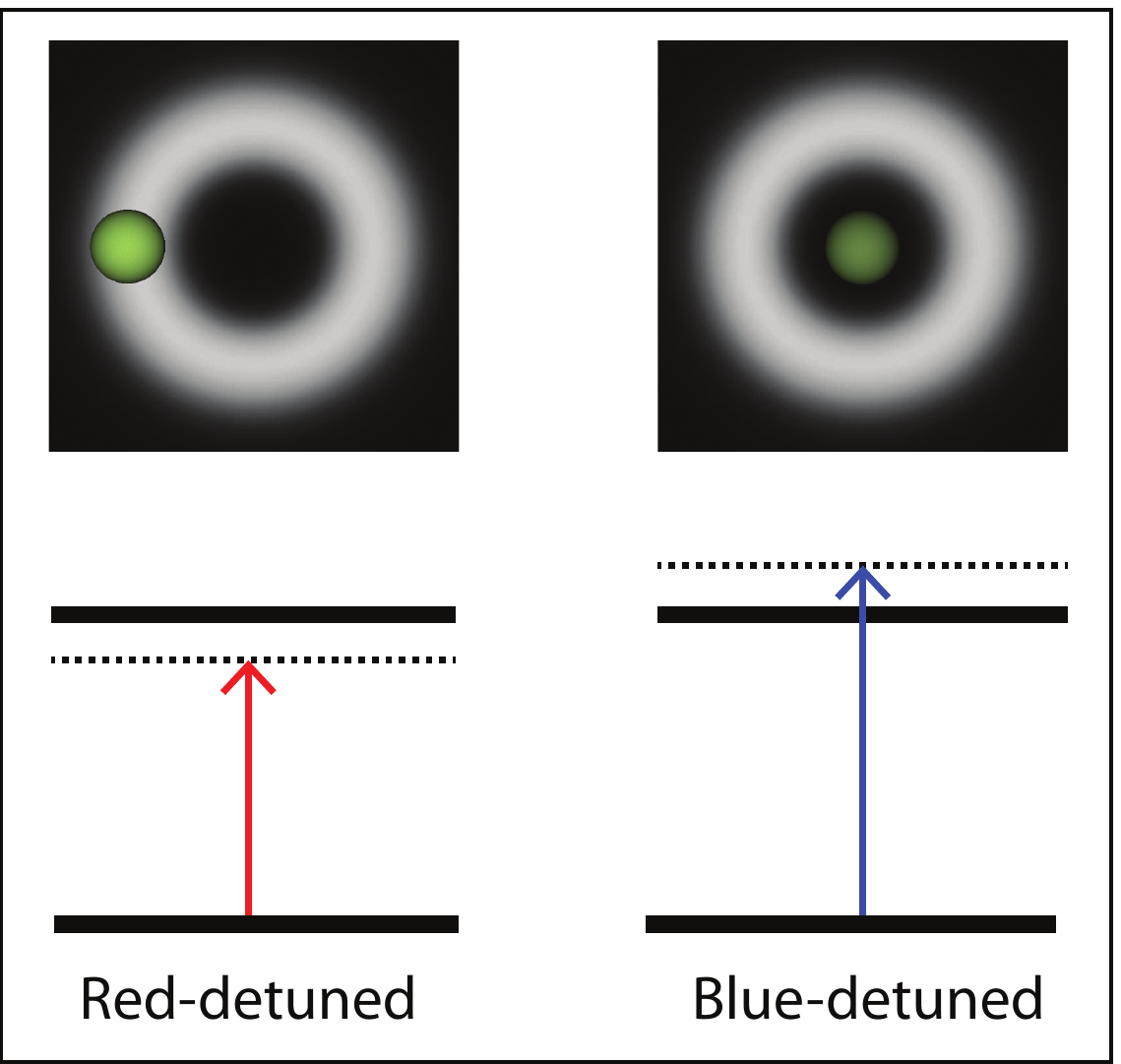}}
\caption{The designs of trapping geometries to confine  ions to regions of maximal or minimal light intensity for red- or blue-detuned LG beam, respectively.}
\end{figure*}
 
In this paper, we develop a theory to calculate the dynamic dipole polarizability of an atomic state with circularly polarized non-paraxial LG beam and apply this to determine magic wavelengths of the transitions $5s_{1/2} \rightarrow 4d_{3/2, 5/2}$ of Sr$^+$ ion. We show that how the OAM and SAM of a focused LG beam affect the dipole polarizability of an atomic state. The coupling of these two kinds of angular momentum increases with the focusing angle.  The impact of focusing angle on the dipole polarizability and magic wavelengths will be interesting to the experimentalists, and we quantify this with our numerical calculations. We also found a number of magic wavelengths for which the ion can be confined  to the nodes  (blue-detuned) or antinodes (red-detuned) of the LG beam.

\section{THEORETICAL FRAMEWORK}
 The  second-order energy shift of  an atom or ion placed in an external oscillating electric field $E(\omega)$ can be estimated from the time-independent perturbation theory as \cite{Mitroy2010} $ \Delta F(\omega)= -\frac{1}{2}\alpha(\omega)E^2$, where $\alpha (\omega)$ is the polarizability of the atomic or ionic energy state at frequency $ \omega $ and $E$ is the magnitude of the external electric field. For  monovalent atomic system with a valence electron in the $v$th orbital, the polarizability can be represented as
\begin{equation}\label{1}
\alpha(\omega)=\alpha_c(\omega)+\alpha_{vc}(\omega)+\alpha_v(\omega)
\end{equation}
 Here, $\alpha_c(\omega)$ and $\alpha_v(\omega)$ are frequency dependent  core polarizability of the ionic core  (in the absence of the valence electron) and valence polarizability of the single valence system, respectively.   $ \alpha_{vc}(\omega) $ represents the correction \cite{Safronova2011} in core polarizability due to presence of the valence electron and is  considered $\omega$ independent due to tightly  bound core electrons.  The core polarizability of an atomic or ionic system can be estimated as \cite{Mitroy2010, Ghosh1993, Dutta2015}
\begin{equation}\label{2}
\alpha_c(\omega)=\frac{2}{3}\sum_{ap}\frac{|\langle \Phi_a||D_\textrm{DF}||\Phi_p\rangle \langle \Phi_a||D_\textrm{RMBPT(2)}||\Phi_p\rangle|(\epsilon_p-\epsilon_a)}{(\epsilon_p-\epsilon_a)^2-\omega^2}.
\end{equation}

Here $a$  and $p$ represent all the core (occupied by electron)  and  virtual orbitals (unoccupied by electron), respectively.  $\langle \Phi_a||D_\textrm{DF}||\Phi_p\rangle$ and   $\langle \Phi_a||D_\textrm{RMBPT(2)}||\Phi_p\rangle$ are reduced dipole matrix elements at the Dirac-Fock (DF)  and the second-order relativistic many-body perturbation theory (RMBPT(2)) levels, respectively. 

To calculate the valence polarizability ($\alpha_v(\omega) $)
of a monovalent system, we consider that a non-paraxial LG beam interacts with cold Sr$^+$ whose de Broglie wavelength is large enough to feel the intensity variation of the focused LG  beam.  Here, the non-paraxial beam is created from a circularly polarized  LG beam by passing it through an objective (lens) with high numerical aperture \cite{Bhowmik2016}. The spot size of the paraxial LG beam is such that it overfills  the entrance aperture radius  of the objective to take full advantage of the high numerical aperture. Because of focusing and the diffraction from the edges of the objective, the SAM and OAM of the light get coupled  and compose into a superposition of plane waves having an infinite  number of
spatial harmonics \cite{Richards1959, Boivin1965}.  Here, we should mention that whenever we refer SAM or OAM, it should
be understood that we mean the corresponding angular momentum of the paraxial LG beam before passing  through the objective lens.
For non-paraxial circularly polarized LG beam,  the electric field  in the laboratory coordinate system can be expressed   as \cite{Bhowmik2016}

\begin{eqnarray}\label{3}
\textbf{E}&=&E_0 e^{-i\omega t}\Bigl[I_0^{(l)}(r _\bot ^\prime ,z ^\prime)e^{il\Phi^\prime}\{\boldsymbol{\hat{\textbf{x}}}(-i)^{l+1}+ \boldsymbol{{\hat{\textbf{y}}}} \beta(-i)^l\} \nonumber \\
&+& I_{2\beta}^{(l)}(r _\bot ^\prime ,z ^\prime)e^{i(l+2\beta)\Phi^\prime}\{\boldsymbol{\hat{\textbf{x}}}(-i)^{l+1}  -\boldsymbol{{\hat{\textbf{y}}}}\beta(-i)^l\}\, \nonumber\\
&-&(2\beta) (-i)^{l} I_\beta^{(l)}(r _\bot ^\prime ,z ^\prime)e^{i(l+\beta)\Phi^\prime}\boldsymbol{\hat{\textbf{z}}} \Bigr]+ c.c.,
\end{eqnarray}
where $\beta$ and $\omega$ are the polarization and frequency of light, respectively.   The amplitude of the focused electric field is $E_0=\dfrac{\pi f}{\lambda} T E_{inc}$, where $E_{inc}$ is the amplitude of incident electric field,  $T$ is the  transmission amplitude of the objective,  and $f$ is its focal length related with $r^\prime$ by $r^\prime=f \sin\theta$ (Abbe sine condition). The coefficient $I_m^{(l)}$, where $m$ takes the values 0, $\pm1$, $\pm2$   in the above expressions, depends on focusing angle ($\theta_{max}$) by \cite{Zhao2007, Bhowmik2016}

\begin{equation}\label{4}
I_m^{(l)}(r _\bot ^\prime ,z ^\prime)=\int_0^{\theta_{max}}d\theta\left({\dfrac{\sqrt{2}r_\bot^\prime }{w_0 \sin\theta}}\right)^{\lvert l \rvert}{(\sin\theta)}^{\lvert l \rvert +1} 
\sqrt{\cos\theta}g_{\lvert m \rvert}(\theta) J_{l+m}(kr_\bot^\prime \sin\theta)e^{ikz^\prime \cos\theta},
\end{equation} 
where $r_\bot^\prime$ is the projection of \textbf{r$^\prime$} on the $xy$ plane, $w_0$ is the waist of the paraxial beam and $J_{l+m}(kr_\bot^\prime \sin\theta)$ is cylindrical Bessel function. The angular functions are  $g_0 (\theta)=1+\cos\theta$, $g_1 (\theta)=\sin\theta$, $g_2 (\theta)=1-\cos\theta$. We consider the incident beam has  circular polarization with  $\beta = \pm 1$.  Therefore, Eq.~(\ref{3}) becomes  

\begin{equation}\label{5}
\textbf{E}=E_0 e^{-i\omega t}\Bigl[\sqrt{2}(-i)^{l+1}I_0^{(l)}e^{il\Phi^\prime}\boldsymbol{\hat{\varepsilon}_{\beta}} 
+ \sqrt{2}(-i)^{l+1} I_{\pm 2}^{(l)}e^{i(l\pm 2)\Phi^\prime}\boldsymbol{\hat{\varepsilon}_{-\beta}} 
\mp 2 (-i)^{l} I_{\pm 1}^{(l)}e^{i(l\pm 1)\Phi^\prime}\boldsymbol{\hat{\textbf{z}}} \Bigr]+ c.c.,
\end{equation}

where the polarization vector $\boldsymbol{\hat{\varepsilon}_{\beta}}=\frac{\boldsymbol{\hat{\textbf{x}}}+i\beta\boldsymbol{\hat{\textbf{y}}}}{\sqrt{2}}$. To make the equation simpler, we have written $I_m^{(l)}(r _\bot ^\prime ,z ^\prime)$ as  $I_m^{(l)}$.     Since, focusing of the LG beam has created three types of local polariztion (right circular, left circular and linear) \cite{Bhowmik2016}, therefore, to conserve  the total angular momentum in each of the parts of  Eq.~(\ref{5}),     OAM should be modiefied accordingly.   Hence,  $\alpha_v(\omega) $ should have the cumulative effect of all three polarized parts of the electric field. Now,  using Eq. ~(\ref{5}), $\alpha_v(\omega)$  will take the form as
\begin{eqnarray}\label{6}
\alpha_v(\omega) =2A_0 \alpha_v^0(\omega)
+ 2\times\left(\frac{m_J}{2J_v}\right) A_1\alpha_v^1(\omega)  
&+& 2 \times \left(\frac{3m_J^2-J_v(J_v+1)}{2J_v(2J_v-1)}\right)A_2\alpha_v^2(\omega),
\end{eqnarray}
where $J_v$ is the total angular momentum of the state $\psi_v$ and $m_j$  is  magnetic component. The parameters $A_i$s are defined as
 $A_0=\left[\{I_0^{(l)}\}^2+\{I_{\pm 2}^{(l)}\}^2+2 \{I_{\pm 1}^{(l)}\}^2\right]$,
 $A_1=\left[\pm \{I_0^{(l)}\}^2 \mp \{I_{\pm 2}^{(l)}\}^2\right]$
 and $A_2= \left[\{I_0^{(l)}\}^2+\{I_{\pm 2}^{(l)}\}^2-2 \{I_{\pm 1}^{(l)}\}^2\right]$. 
$\alpha_v^0(\omega)  $, $ \alpha_v^1(\omega) $ and $ \alpha_v^2(\omega) $ are the scalar, vector and tensor parts, respectively, of valence polarization and can be written as \cite{Mitroy2010, Dutta2015} 
 \begin{equation}\label{7}
 \alpha_v^0(\omega)=\frac{2}{3(2J_v+1)}\sum_n \frac{|\langle\psi_v||d||\psi_n\rangle|^2\times(\epsilon_n-\epsilon_v)}{(\epsilon_n-\epsilon_v)^2-\omega^2},
 \end{equation}
\begin{equation}\label{8}
\alpha_v^1(\omega)=-\sqrt{\frac{6J_v}{(J_v+1)(2J_v+1)}}\sum_n (-1)^{J_n+J_v} \left\{\begin{array}{ccc} J_v & 1 & J_v\\ 1 & J_n& 1 \end{array}\right \}\frac{|\langle\psi_v||d||\psi_n\rangle|^2 \times 2\omega}{(\epsilon_n-\epsilon_v)^2-\omega^2},
\end{equation}
and
\begin{equation}\label{9}
\alpha_v^2(\omega)=4\sqrt{\frac{5J_v(2J_v-1)}{6(J_v+1)(2J_v+1)(2J_v+3)}}\sum_n (-1)^{J_n+J_v} \left\{\begin{array}{ccc} J_v & 1 & J_n\\ 1 & J_v& 2 \end{array}\right \} \frac{|\langle\psi_v||d||\psi_n\rangle|^2\times(\epsilon_n-\epsilon_v)}{(\epsilon_n-\epsilon_v)^2-\omega^2}.
\end{equation}
Therefore, $ \alpha_v(\omega)$ directly  depends on different combinations of integrals $ I_m^{(l)}$. And, these integrals  can be modified by changing the combination of SAM and OAM of the incident  LG beam and  numerical aperture of the objective.  Therefore the polarizability   can also be tuned  with the focusing angle of the non-paraxial LG beam.

\section{NUMERICAL RESULTS AND INTERPRETATIONS}

 The major aim of this work is  to calculate  precise values of magic wavelengths associated with the $5s_{1/2}\rightarrow 4d_{3/2,5/2}$
 transitions of Sr$^+$  ion. Therefore as stated earlier,  we need to estimate dynamic polarizabilities of the $5s_{1/2}$, $4d_{3/2}$, and $4d_{5/2}$ states of this ion for different magnetic sublevels.  Using Eqs ~(\ref{7}), ~(\ref{8}), and ~(\ref{9}), one can calculate the scalar, vector and tensor parts of the valence polarizabilities, respectively, for the associated valence configurations of these states. The precise estimations of  $E1$ transition amplitudes and corresponding transition energies highlight the accuracy of our calculations.  In order to evaluate these quantities, we use a relativistic coupled cluster (RCC) theory having wave operators associated with single and double and valence triple excitations in linear and non-linear forms. This is similar in the spirit of CCSD(T) \cite{Raghavachari1989} used by many quantum chemists. Our RCC wavefunctions,  based on the corresponding DF wavefunctions,  produce highly precise \textit{E1} transition amplitudes as discussed in our earlier work \cite{Dutta2013, Dutta2016, Roy2014, Bhowmik2017a, Bhowmik2017b, Das2018}.
 
Table I presents  a comparison of the most important  reduced dipole matrix elements as calculated by us with the corresponding theoretical results of Safronova \cite{Safronova2010} and  some experimental measurements \cite{Pinnington1995, Gallagher1967}.   Safronova estimated  the results by using an all-order single-double with partial triple (SDpT)
excitations method in linearized approximation. The small difference between the results coming from the addition of some nonlinear terms in our present theory. Also, Safronova used B-spline bases to construct the Dirac-Fock orbitals,   whereas we  consider Gaussian-type orbital (GTO)  bases to generate these orbitals. The table also includes a comparison of the wavelengths of the transitions calculated by our RCC method with the corresponding wavelengths as obtained from the  National Institute of Standards and Technology (NIST) \cite{NIST}.

\begin{threeparttable}[!h]
  \caption{Calculated $E1$ transition amplitudes (in a.u.) and their comparison with other results (Others). The experimental ($\lambda_{\textrm{NIST}}$) and RCC($\lambda_{\textrm{RCC}}$) wavelengths are presented in \AA. "Others" are calculated using excitation  energies from NIST \cite{NIST}  and the oscillator strengths presented in the references.}
\centering
\begin{tabular}{cccccc l}

\hline \hline

   Transition &$\lambda_{\textrm{RCC}}$  & $\lambda_{\textrm{NIST}}$   & RCC &  & & Others
           \\ [0.2ex]
\hline

$5s_{1/2}  $  $\rightarrow$ $5p_{1/2}$  &4191.32

&4216.88& 3.1062

   & & & 3.0903$^a$, 3.12$^b$  \\ 

 \hspace{0.9cm}   $\rightarrow$ $5p_{3/2}$  &4053.71

&4079.05& 4.38971
  & & & 4.3704$^a$, 4.40$^b$ \\ 
 
   $4d_{3/2}  $  $\rightarrow$ $5p_{1/2}$ &11439.88
&10918.61
 
& 3.08262 &  & & 3.1113$^a$,  3.47(32)$^c$ \\ 

    \hspace{0.9cm}$\rightarrow$ $5p_{3/2}$&10469.81
 &10040.18
 & 1.36854 &  & & 1.3820$^a$,  1.45(14)$^c$ \\ 

   \hspace{0.9cm}  $\rightarrow$ $4f_{5/2}$&2172.71
 &2153.56

 & 2.82947 &  & & 2.9172$^a$ \\
   
  $4d_{5/2}  $  $\rightarrow$ $5p_{3/2}$ &10807.36
 &10329.25

& 4.1498 &  & & 4.1833$^a$ \\ 
 
  \hspace{0.9cm}  $\rightarrow$ $4f_{5/2}$&2186.88
 &2166.57

 & 0.76694 &  & & 0.7887$^a$ \\
   
    \hspace{0.9cm}  $\rightarrow$ $4f_{7/2}$ &2186.95
&2166.59

 & 3.43082 & & & 3.5214$^a$ \\
  
\hline  
\label{table:nonlin} 
\label{II}
\end{tabular}
\vspace*{-1.0cm}
    \begin{tablenotes}
    \begin{tiny}
\item $a \rightarrow$ (Theoretical) \cite{Safronova2010},
$b \rightarrow$ (Experimental) \cite{Pinnington1995},
$c \rightarrow$ (Experimental) \cite{Gallagher1967}
\end{tiny}
\end{tablenotes}
\end{threeparttable}
 
\vspace*{0.5cm}
 
To calculate the required polarizabilities for $5s_{\frac{1}{2}}$, $4d_{\frac{3}{2}}$ and $ 4d_{\frac{5}{2} }$ states, we use Eq.~(\ref{1}), Eq.~(\ref{2}), Eq.~(\ref{6}), Eq.~(\ref{7}), Eq.~(\ref{8}), and Eq.~(\ref{9}).  The ionic core polarizability ($\alpha_c (\omega)$) is irrespective of the position of valence electron and can be calculated quite accurately using Eq.~(\ref{2}). The valence polarizability needs the most important  attention as it can be affected significantly by the electron-correlation due to loser binding of a valence electron to the nucleus. The  $E1$ matrix elements present in the  valence polarizability expressions (Eq.~(\ref{7}), Eq.~(\ref{8}), and Eq.~(\ref{9})) are considered at different levels of theoretical considerations depending on their significance to the sum.   The most dominant and therefore important contributions to the valence polarizabilities come from the parts of the sums in Eq.~(\ref{7}), Eq.~(\ref{8}), and Eq.~(\ref{9}) which are involved with   the intermediate states in the ranges of $5^2P-8^2P$ and $4^2F-6^2F$. Therefore, the $E1$ matrix elements associated with these intermediate states are calculated using the correlation exhaustive RCC method. RMBPT(2) \cite{Johnson1996}, which includes core polarization correction on top of the DF approximation,  is used to calculate  the comparatively less significant $E1$ matrix elements in the polarizability expressions with intermediate states from  $9^2P-12^2P$  and $7^2F-12^2F$.  The intermediate states with $ n=13$ to $ 25 $ in Eq.~(\ref{7}), Eq.~(\ref{8}), and Eq.~(\ref{9})  contribute by a small value, and therefore, without loss of significant accuracy to the polarizability value, they are computed using the DF wavefunctions. For $n$ greater than 25, the sums are expected to contribute by a very little amount and thus  are neglected. To obtain  better accuracy in calculating a total polarizability value, we have used the experimental  transition energies \cite{NIST}  to calculate the most dominant part of corresponding valence polarizability. 

In Table II, we compare   static values of  valence scalar and tensor polarizabilities for the $5s_{\frac{1}{2}}$ and $4d_{\frac{3}{2}, \frac{5}{2}}$ states with the  theoretical \cite{Safronova2010, Kaur2015, Jiang2009} and experimental \cite{Barklem2000} values as available in the literature. Both Jiang \textit{et al.} \cite{Jiang2009} and Safronova \cite{Safronova2010} adopted almost a similar strategy in calculations of the most dominant contributor to their total polarizability values, i.e.,  the valence polarizabilities.  They used approximately similar kind of all-order relativistic many-body perturbation method where single, double and partial triple excitations  in this method are considered in linear form. As mentioned earlier, the present approach accounts  these excitations in both linear and non-linear forms. Also, both of them applied random phase approximation (RPA) to calculate the core polarizability, whereas, we use RMBPT(2) to estimate the core polarizability value.     Kaur \textit{et al.} applied  a relativistic coupled-cluster method with single and double excitations to compute dominant portions of the valence polarizabilities of the ground and excited states \cite{Kaur2015}. However, they used the same RPA approximation as used by Jiang \textit{et al.} and Safronova to calculate the core polarizability.  All the  theoretical calculations are in very close agreement for core and core-valence parts of the corresponding total polarizabilities. The computed static core polarizability ($\alpha_c(0)$) of the ion is 6.103 a.u., and the  static core-valence parts of the polarizabilities ($\alpha_{vc}(0)$) for the states $5s_{\frac{1}{2}}$, $4d_{\frac{3}{2}}$ and  $4d_{\frac{5}{2}}$ are  $-0.25$ a.u., $-0.38$ a.u. and $-0.42$ a.u., respectively. In Table II, we compare only valence parts of the corresponding static polarizabilities among the estimations from different calculations.   The  experimental measurement of Barklem and O'Mara \cite{Barklem2000} shows a difference of the valence  polarizability of $5s_{\frac{1}{2}}$ state by around 0.2\%.  Nevertheless, the  agreements of our calculated  energies, amplitudes of $E1$ transitions, static values of scalar and tensor polarizabilities with the estimations by other theoretical and experimental groups can indicate a  good  calibration of our present calculations. We can claim now that our present approach of calculating dynamic polarizability is accurate enough to study the effect  of focused  LG beam on Sr$^+$ in terms of magic wavelengths.

\begin{threeparttable}[!h]
  \caption{ Static valence scalar ($\alpha_v^0$)  and static tensor ($\alpha_v^2$) polarizabilities (in a.u.) for the states $5s_{1/2}$, $4d_{3/2}$ and  $4d_{5/2}$ and their comparisons with the other results (Others).}
\centering
\begin{tabular}{cccclcccl}

\hline 
&&\multicolumn{3}{c}{\textbf{$\alpha_v^0$}} && \multicolumn{3}{c}{\textbf{$\alpha_v^2$}}\\  
\cline{3-5}\cline{7-9}
  State & &Present   &   & Others& & Present & &Others
           \\ [0.2ex]
\hline

$5s_{1/2}  $ && 87.68 && 86.374$^a$,  85.75$^b$ &  &  &  &  \\ 
                    &&          && 85.75$^c$,  87.5$^d$     &  &  &   &  \\ 
                   
$4d_{3/2}$  && 55.92 && 57.87$^a$, 58.78$^b$    & & $-34.67$
& & $-35.50(6)^a$,  $-35.26^b$ \\ 
                   &&          && 51.20$^d$                       &  &   &  &  \\ 
$4d_{5/2}$  && 56.21 && 56.618$^a$, 57.11$^b$  & & $-47.12$ & & $-47.70(8)^a$, $-47.35^b$ \\ 
 
                   &&         &&  56.59$^c$, 51.20$^d$  &&  && $-47.70(3)^c$ \\ 
\hline 
\label{table:nonlin} 
\label{II}

\end{tabular}

\begin{tablenotes}
\begin{tiny}
\vspace{-0.8cm}
\item  $a\rightarrow$ Safronova \cite{Safronova2010},
 $b\rightarrow $ Kaur \textit{et al.}  \cite{Kaur2015},
 $c\rightarrow$  Jiang \textit{et al.} \cite{Jiang2009},\\
 \vspace{-.8cm}
 \item $d\rightarrow $ Barklem and O'Mara  \cite{Barklem2000}
\end{tiny}
\end{tablenotes}
\end{threeparttable}

For the sake of comparison with LG beam, we have first plotted the dynamical  polarizability of $5s_{\frac{1}{2}}$ and $4d_{\frac{3}{2}, \frac{5}{2}}$ states of Sr$^+$ with the frequency of circularly polarized Gaussian light in FIG. 2 and 3. This is extremely important for many high precision experiments in quantum optics, especially, identifications of magic wavelengths corresponds to high polarization.   Point to be noted from the two figures that variation path of polarizabilities exchanged between states with $(J,m_J)$ and $(J,-m_J)$ by changing the direction of polarization of light, as expected. However, only one estimation is found in literature \cite{Kaur2015}  with one of the possible magic  wave numbers which is not even for high polarizability value. The analysis in the broad range of magic frequencies including the one, corresponding highest polarizability presented in FIG 2 and FIG. 3.   The actual values of the magic wavelengths and corresponding polarizabilities due to circular and linearly polarized light are tabulated in TABLE III. One can see that there are multiple  magic wavelengths found in the presented frequency range for  the transitions between the magnetic sublevels of $5s_{\frac{1}{2}}$ and magnetic sublevels of $4d_{\frac{3}{2},\frac{5}{2}}$ states.     The tabulated magic wavelengths fall in the near-infrared, visible and UV  regions of the frequency spectrum. All the magic wavelengths, which belong to the visible and UV  regions,  favour blue-detuned trapping and which belong to the  near-infrared region, support red-detuned trap scheme. In some cases, no magic wavelength is found for a particular range of the spectrum, and we kept them the slot as blank in the TABLE II and tables henceforth.

\begin{table}[h]
\scriptsize
  \caption{Magic wavelengths (in nm)  of Sr$^+$ for linearly and circularly polarized Gaussian light for the transitions $5s_{1/2}(+1/2)  $  $\rightarrow$ $4d_{3/2, 5/2}(m_J)  $.}
\centering
\begin{tabular}{cccccc|cccccc|ccc}

\hline \hline
     
\\
\multicolumn{6}{c}{\textbf{Circularly polarized  (SAM=+1)}}&\multicolumn{6}{c}{\textbf{Circularly polarized  (SAM=$-1$)}} & \multicolumn{3}{c}{\textbf{Linearly polarized }}\\  
  \hline 
   State     & $\lambda_{\textrm{magic}}$& $\alpha$ &  State     & $\lambda_{\textrm{magic}}$& $\alpha$&  State      & $\lambda_{\textrm{magic}}$& $\alpha$ &  State     & $\lambda_{\textrm{magic}}$& $\alpha$&  State      & $\lambda_{\textrm{magic}}$& $\alpha$
           \\ [0.2ex]

 $(J, m_J)$   & &  &   $ (J, m_J)$     & & &  $ (J, m_J)$    & &  &   $(J, m_J)$     & & &  $(J, m_J)$     & & 
           \\ [0.2ex]
   \hline 
$(\frac{3}{2},+\frac{1}{2})  $&1062.08	&	107.96&$(\frac{3}{2},-\frac{1}{2})  $	&	1054.71	&	108.71& 
$(\frac{3}{2},+\frac{1}{2})  $&1052.27	&	110.54
&$(\frac{3}{2},-\frac{1}{2})  $	&	1059.61	&	110.52
&

$(\frac{3}{2},|\frac{1}{2}|)  $&1052.27	&	109.51\\
&	402.15	&	22.06  & & 402.15	&	0.51
& &420.33	&	-2.29&&421.10	&	20.05&&406.82	&	10.96

	\\
&212.81	&	-24.13   &&	212.81	&	-24.13&
&420.33	&	-2.29&&212.91	&	-24.63
&
&
   212.91	&	-24.63
\\

&198.19	&	-19.72	&&	200.28	&	-20.45	&
&200.10	&	-19.59&&197.76	&	-18.56
&&
  198.53	&	-19.65\\

$(\frac{3}{2},+\frac{3}{2})  $&	1072.08	&	107.96	&$(\frac{3}{2},-\frac{3}{2})  $&	1732.45	&	98.34&  $(\frac{3}{2},+\frac{3}{2})  $ &1739.06	&	99.37&$(\frac{3}{2},-\frac{3}{2})  $&1072.08	&	109.28
&
$(\frac{3}{2},|\frac{3}{2}|)  $&	1130.60	&	106.99\\

&	861.31	&	118.16	&&	953.21	&	112.54&& 949.24	&	115.06
&& 869.53	&	120.9
&& 929.86	&	115.06\\

&	402.15	&	46.64	&&	402.15	&	-18.04&& 420.33	&	-19.59
&&421.49	&	48.45
&&	406.82	&	15.22\\
&	212.81	&	-24.13	&&	220.43	&	-27.37	&& 220.01	&	-27.72
&&212.91	&	-24.63
&&213.31	&	-25.54\\

&	198.79	&	-19.98	&&	215.33	&	-25.75	&&215.43	&	-25.66
&&198.19	&	-18.56
&&202.86	&	-20.40\\

$(\frac{5}{2},+\frac{1}{2})  $&	1119.49	&	106.44& $(\frac{5}{2},-\frac{1}{2})  $	&	1119.49	&	106.44&$(\frac{5}{2},+\frac{1}{2})  $&1125.02	&	107.90&$(\frac{5}{2},-\frac{1}{2})  $&1125.02	&	107.90&$(\frac{5}{2},|\frac{1}{2}|)  $ &1119.49	&	107.13\\

&	621.60	&	158.93	&&-&-&& -&-&&619.91	&	167.27
&&
-&-\\
&	585.65	&	174.91	&&-&-&&-&-&&588.67	&	184.45
&&
-&-\\

&	402.15	&	27.79	&&	402.15	&-2.56	&
&420.33	&	-6.41
&&421.49	&	28.18
&&407.18	&	12.15\\

&	212.12	&	-25.27	&&	212.12	&	-25.27	&&	212.02	&	-24.63
&&212.02	&	-24.63
&&212.02	&	-24.82
\\
&	202.68	&	-21.38	&&	198.79	&	-20.10	&&	198.53	&	-18.56
&&202.32	&	-20.62
&&200.72	&	-20.31\\
$(\frac{5}{2},+\frac{3}{2})  $&	1513.73	&	99.57	&$(\frac{5}{2},-\frac{3}{2})  $&	2301.18	&	95.79&$(\frac{5}{2},+\frac{3}{2})  $&2278.17	&	96.48
&$(\frac{5}{2},-\frac{3}{2})  $& 1503.74	&	101.12
& $(\frac{5}{2},|\frac{3}{2}|)  $&1793.83	&	99.14\\

&	1119.49	&	105.67	&&	1119.49	&	106.44&&1125.02	&	107.9
&&1125.02	&	107.9
&&	1119.49	&	107.13\\

&	631.95	&	155.15&	&	700.97	&	137.63	&&698.82	&	143.73
&& 631.07	&	163.83
&&-&-\\
&	556.33	&	194.67	&&	642.64	&	151.29	&&644.46	&	159.28
&& 561.13	&	205.76
&&-&-\\

&	402.15	&	60.94	&&	402.15	&	-32.60	&&420.33	&	-39.86
&&422.27	&	64.72
&&	407.18	&	14.02\\
&	212.12	&	-25.27	&&	212.12	&	-25.27	&&212.02	&	-24.63
&&212.02	&	-24.63
&&	212.02	&	-24.82\\
&	205.98	&	-22.69	&&	194.38	&	-18.37	&&194.13	&	-17.53
&&205.89	&	-21.08
&&	202.05	&	-20.73\\

$(\frac{5}{2},+\frac{5}{2})  $&	637.25	&	152.84	&$(\frac{5}{2},-\frac{5}{2})  $&	839.10	&	120.10&$(\frac{5}{2},+\frac{5}{2})  $&831.45	&	123.88
&$(\frac{5}{2},-\frac{5}{2})  $&635.47	&	161.94
&$(\frac{5}{2},|\frac{5}{2}|)  $&- &-\\

&	526.74	&	222.85	&&	642.64	&	152.06	&&640.83	&	159.71
&&532.90	&	235.05
&&-&-\\

&	402.15	&	94.25	&&	402.15	&	-59.68	&&420.33	&	-68.27
&&421.49	&	104.23
&&	407.18	&	17.66\\
&	212.12	&	-25.27	&&	212.12	&	-25.27	&&	212.02	&	-24.63
&&212.02	&	-24.63
&&212.02	&	-24.82
\\
&	209.58	&	-24.00	&&	187.12	&	-15.79	&&186.66	&	-15.46
&&209.68	&	-22.57
&&	204.96	&	-21.85\\

\\

\hline
\hline
\label{table:nonlin} 
\label{I}

\end{tabular}
\end{table}

\begin{figure*}[!h]
\subfloat[]{\includegraphics[trim = 1cm 3.0cm 0.1cm 3.5cm, scale=.40]{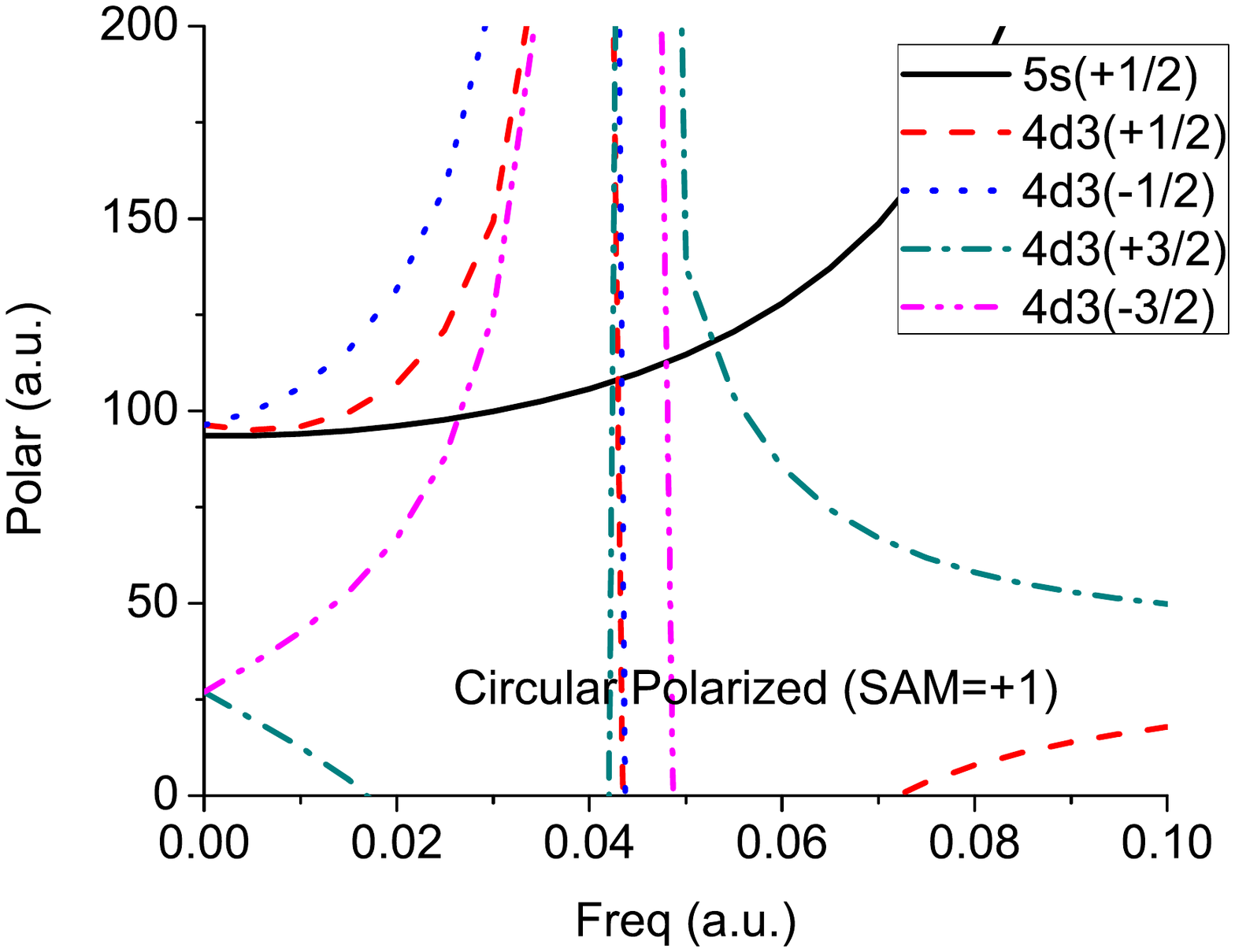}}
\subfloat[]{\includegraphics[trim = 1cm 3.0cm 0.1cm 3.5cm, scale=.40]{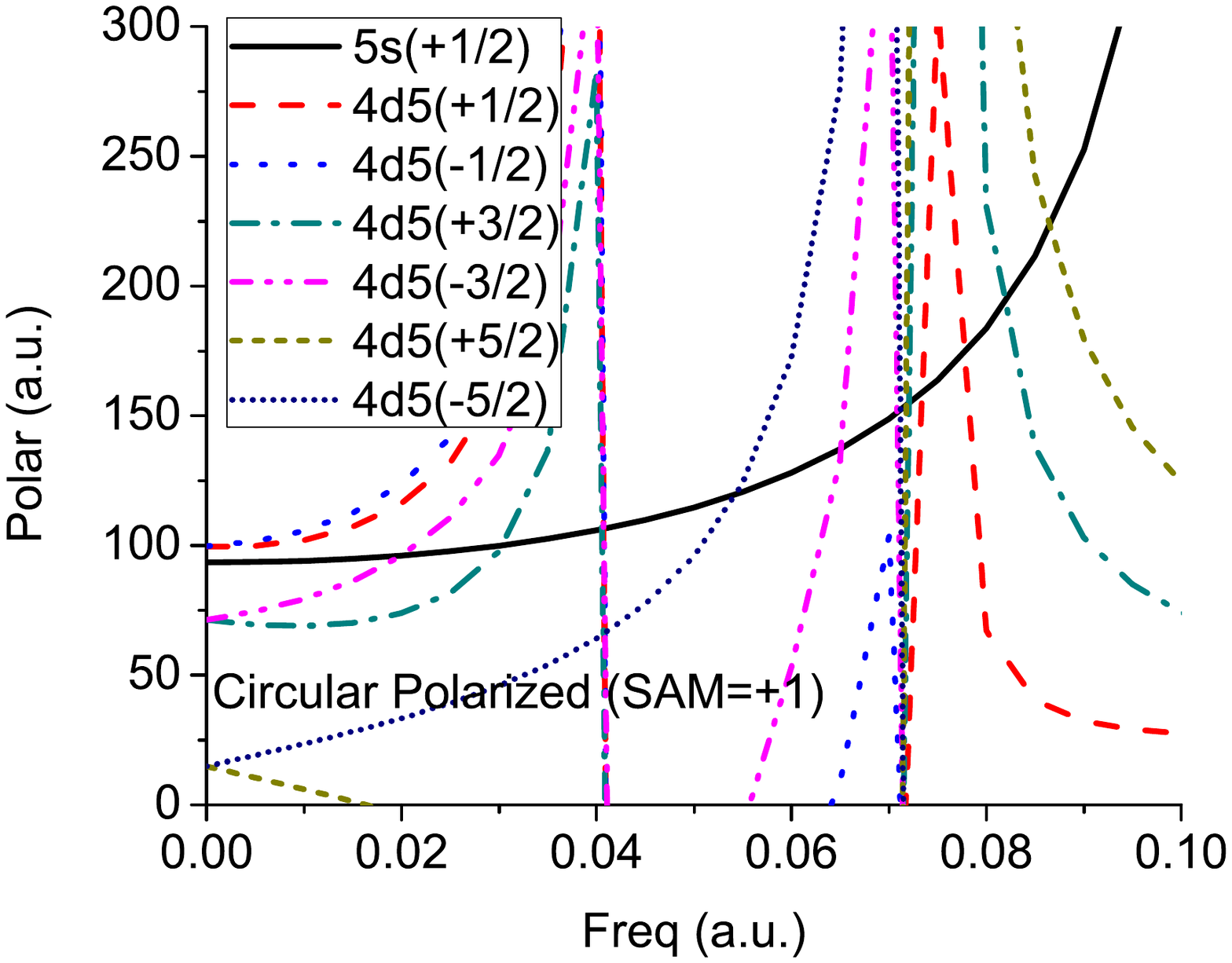}}\\
\subfloat[]{\includegraphics[trim =  1cm 3.0cm 0.1cm 3.1cm,scale=.40]{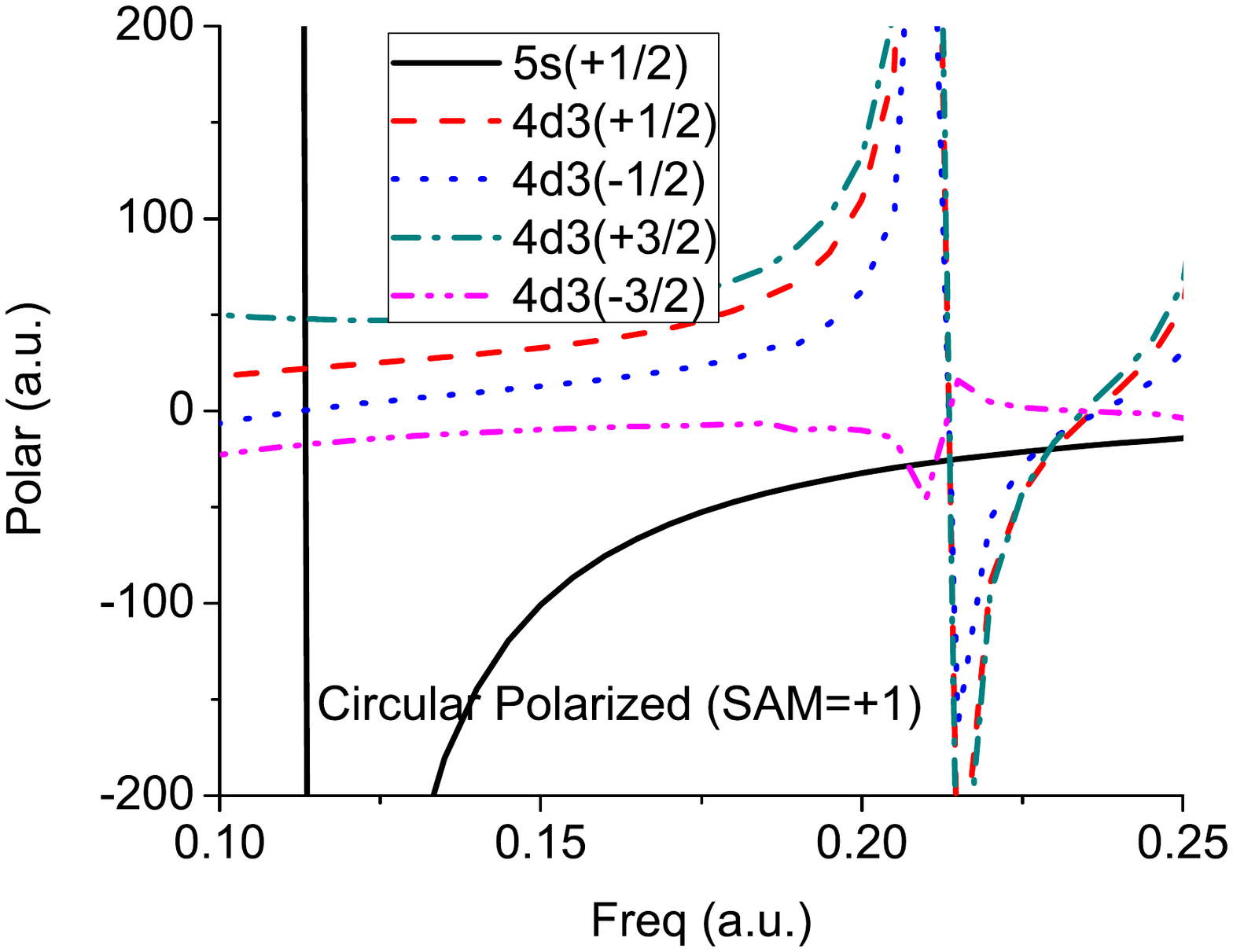}}
\subfloat[]{\includegraphics[trim =  1cm 3.0cm 0.1cm 3.1cm, scale=.40]{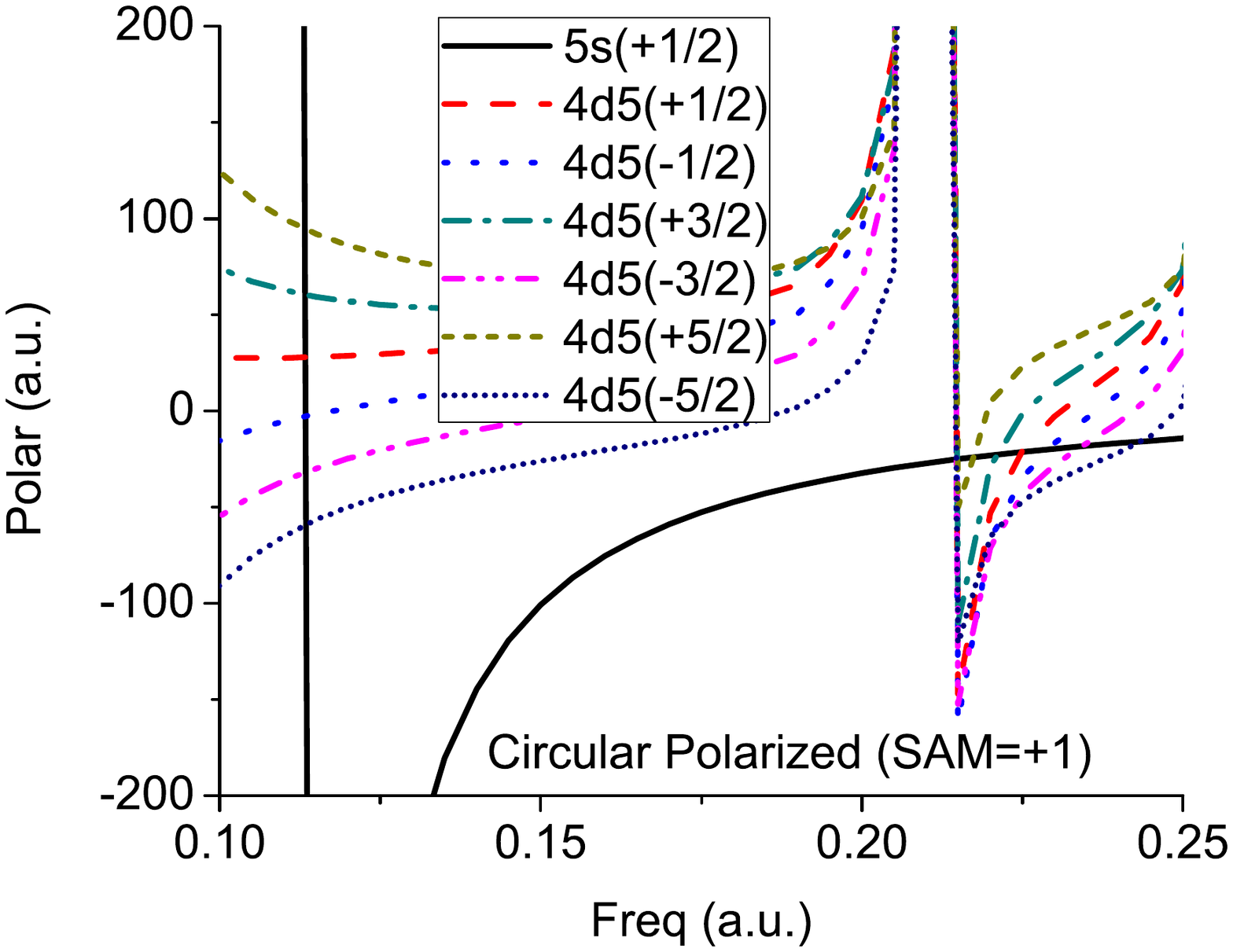}}
\caption{Frequency (Freq) dependence of polarizabilities
(Polar) for the $5s_{\frac{1}{2}}$ and $4d_{\frac{3}{2},\frac{5}{2}}$ states for circular polarized Gaussian light with SAM=+1. The brackets indicate the magnitudes of the magnetic components. Fig. (a) and (c) are for the $5s_{\frac{1}{2}}$ and $4d_{\frac{3}{2}}$ states and  Fig. (b) and (d) are for the $5s_{\frac{1}{2}}$ and $4d_{\frac{5}{2}}$.}
\end{figure*}

\begin{figure*}[!h]
\subfloat[]{\includegraphics[trim = 1cm 3.0cm 0.1cm 3.5cm, scale=.40]{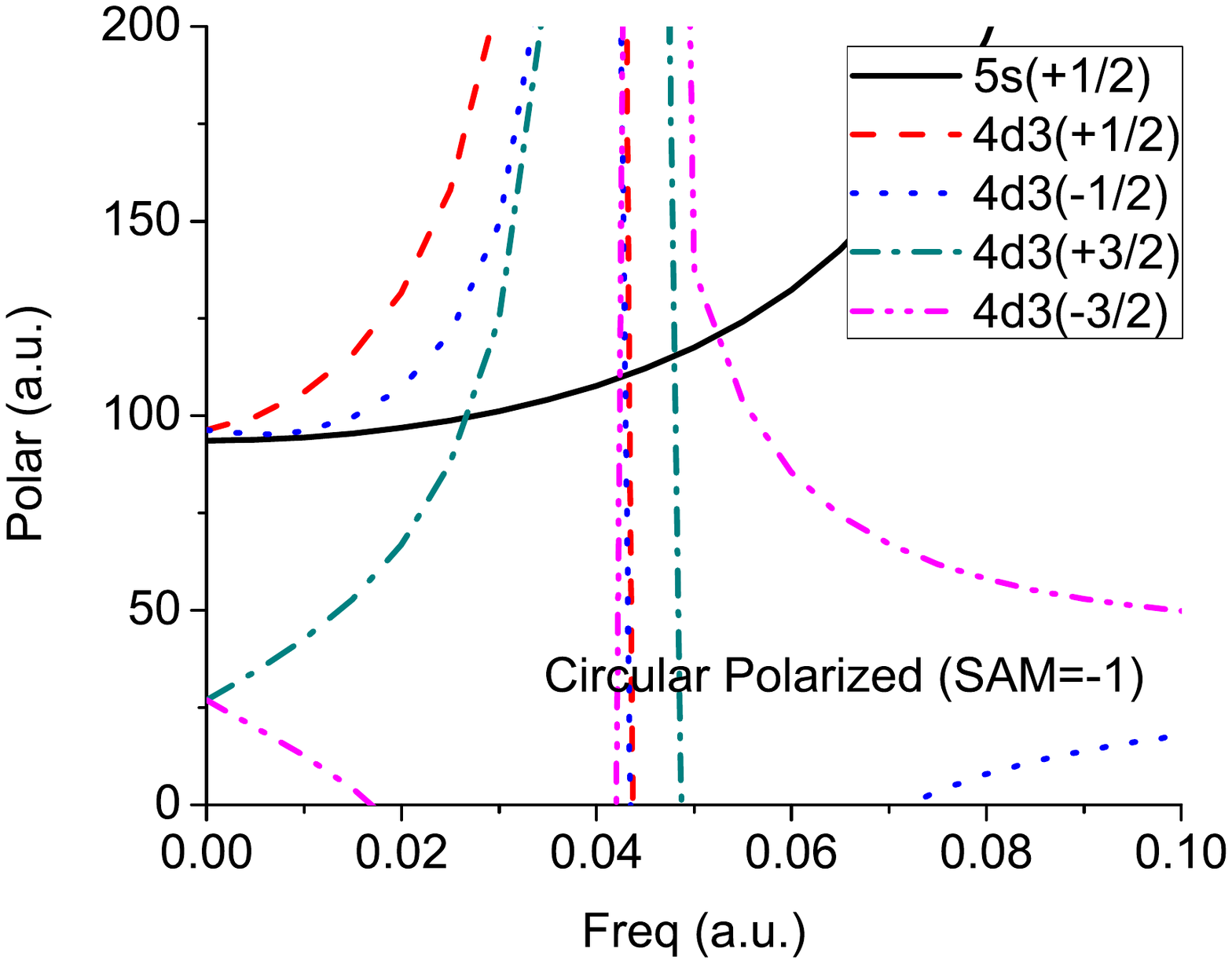}}
\subfloat[]{\includegraphics[trim = 1cm 3.0cm 0.1cm 3.5cm, scale=.40]{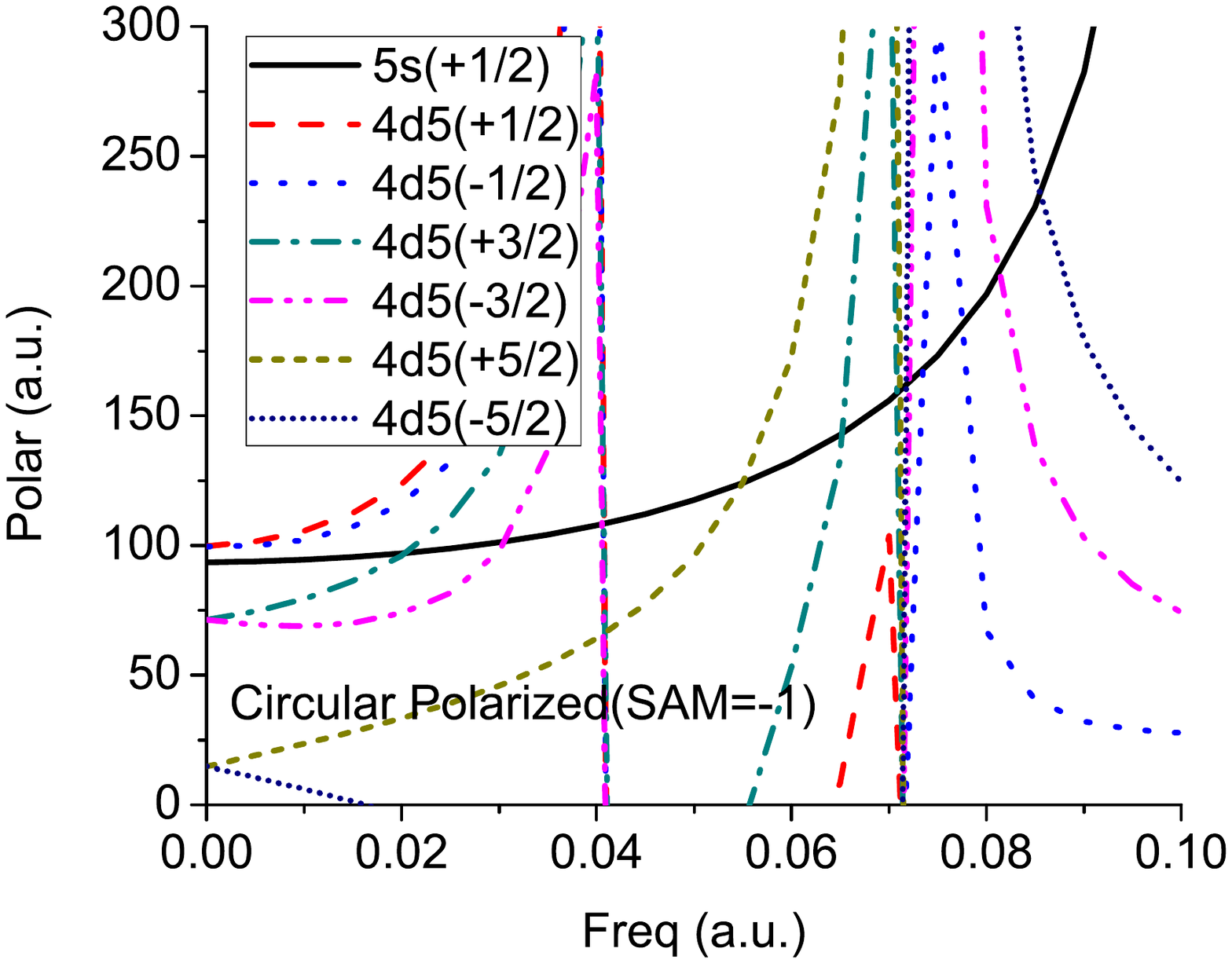}}\\
\subfloat[]{\includegraphics[trim =  1cm 3.0cm 0.1cm 3.1cm,scale=.40]{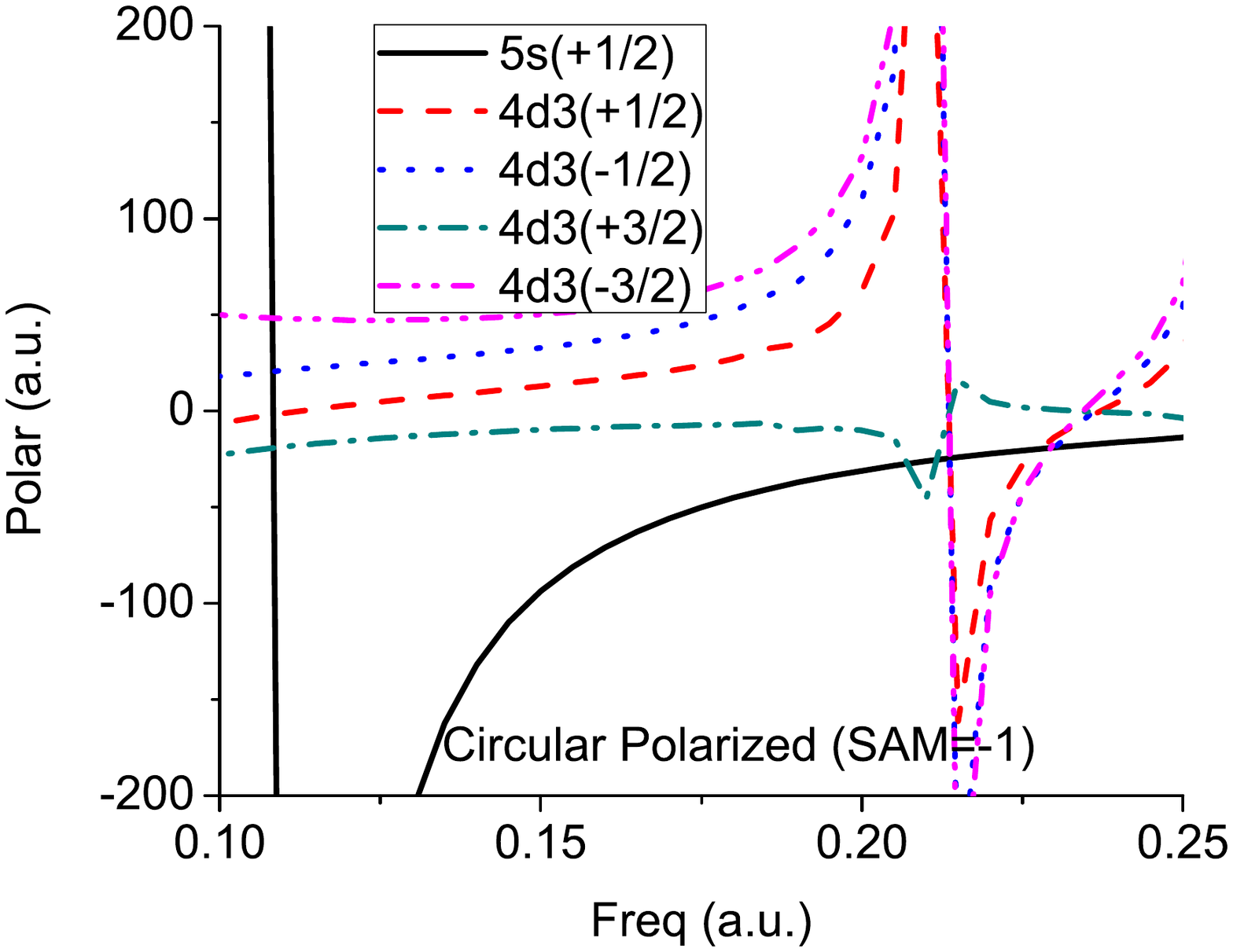}}
\subfloat[]{\includegraphics[trim =  1cm 3.0cm 0.1cm 3.1cm, scale=.40]{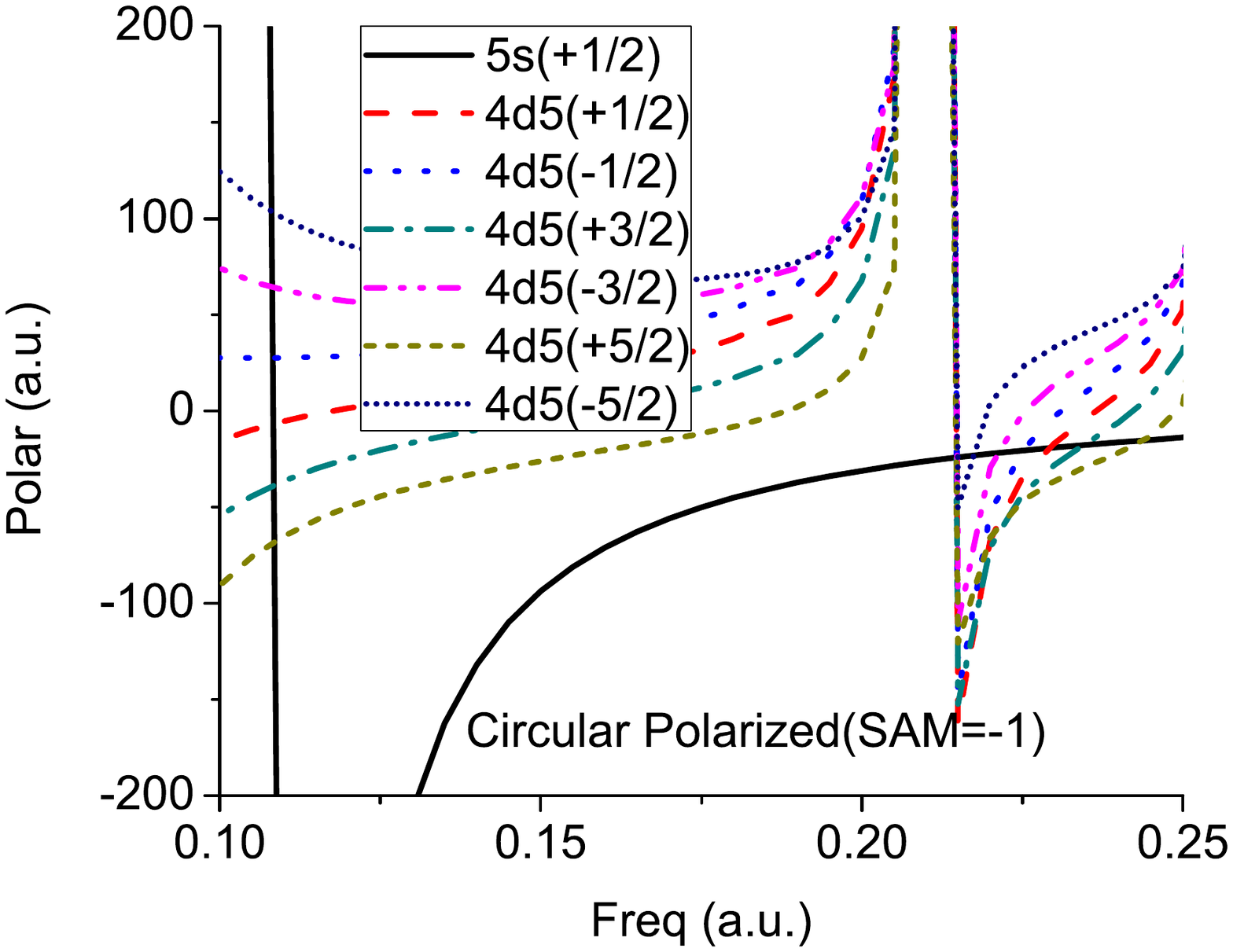}}
\caption{Frequency (Freq) dependence of polarizabilities
(Polar) for the $5s_{\frac{1}{2}}$ and $4d_{\frac{3}{2},\frac{5}{2}}$ states for circular polarized Gaussian light with SAM=$-1$. The brackets indicate the magnitudes of the magnetic components. Fig. (a) and (c) are for the $5s_{\frac{1}{2}}$ and $4d_{\frac{3}{2}}$ states and  Fig. (b) and (d) are for the $5s_{\frac{1}{2}}$ and $4d_{\frac{5}{2}}$.}
\end{figure*}

In FIG.  4 and 5, we present the plots of   the dynamic  polarizabilities of $5s_{\frac{1}{2}}$ and $4d_{\frac{3}{2}, \frac{5}{2}}$ states considering the external field of  non-paraxial LG beam which is focused  at angle of 50$^\circ$.  Here, we have chosen   angular momenta of the incident beam as (OAM, SAM) $= (+1, +1)$ and $(+1, -1)$ to demonstrate their dependencies on the polarizabilities. The maximum polarizabilities observed in these plots correspond to the polarizabilities at the resonance transitions.  For the $5s_{1/2}  $,  $4d_{3/2}  $ and $4d_{5/2}$ states, the resonances occur due to the $5s_{1/2}  \rightarrow 5p_{1/2, 3/2}  $,
$4d_{3/2}  \rightarrow 5p_{1/2, 3/2}  $, and $4d_{5/2}  \rightarrow 5p_{ 3/2}  $ transitions,
respectively, in the chosen spectral range.  These figures show a number of crossings  of the polarizabilities of  $5s_{\frac{1}{2}}$ and multiplets of $4d_{\frac{3}{2}, \frac{5}{2}}$ states. These crossings indicate magic wavelengths of the externally focused LG field for which the transition wavelengths between $5s_{\frac{1}{2}}$ and $4d_{\frac{3}{2}, \frac{5}{2}}$ states are remain unaffected. There have been remarkable difference  of polarizability observed  compare to Gaussian beam at low frequency region.
Significant changes observed for interaction with right circularly LG beam and found extra magic wavelength for $(J,m_J)= (3/2,\pm 1/2)$ in the near-infrared region. Other changes are clear from the comparisons of the polarization plots presented in FIG. 4-5 with FIG. 2-3 and obvious from magic wavelengths tabulated in TABLE IV, V and VI  for different focusing angles ( 50$^\circ$, 60$^\circ$ and 70$^\circ$) and the combination of OAM and SAM ( $(+1, +1)$, $(+1, -1)$, $(+2, +1)$ and $(+2, -1)$ ) with TABLE III.  Since the OAM and SAM are coupled in  case of non-paraxial  LG beam and more  focusing yields more  stronger coupling, from our previous experience with focused LG beam \cite{Bhowmik2016}, we expect the changes of values of  magic wavelengths should appreciable for the chosen large focusing angles of the beam.  The presented magic wavelengths in the tables span from the infrared to ultraviolet (UV) region in the energy spectrum. It is interesting to note that the effects of focusing angles are appreciable for larger magic wavelengths. Also, for those wavelengths,  the values of magic wavelengths are reducing significantly with  the increase of focusing angles, but the polarizabilities are increasing. Our study shows that the change of polarization with the focusing angle is very small for $5s_{\frac{1}{2}}$ state. Therefore, the changes of magic wavelengths seen in the tables are mostly due to the variation of polarizabilities of  $4d_{\frac{3}{2}, \frac{5}{2}}$ states. This phenomenon is significant for trapping of atoms or ions, as magic wavelengths with large polarizabilities will be more helpful to experimentalists for the trapping. In the range of our chosen spectrum,  we find a set of five magic wavelengths for $5s_{\frac{1}{2}}$ to $4d_{\frac{3}{2}, \frac{5}{2}}$ transitions for all the considered combinations of OAM, SAM and focusing angles apart from  few cases.

The Table IV, V and VI show that the infrared or near-infrared magic wavelengths   region of the energy spectrum, have larger values of  polarizabilities compare to the visible and UV regions. Therefore the magic wavelengths  in infrared or near-infrared region are highly recommended for trapping using a red-detuned trap (see FIG. 1), where the ions can be trapped in the region of high intensity of the LG beam \cite{Kennedy2014, Arlt2001}. Also, the magic wavelengths and the corresponding polarizabilities at these regions  show significant variation with focusing angle and choice of  initial OAM and SAM compare to the same at the visible and UV region. Otherway to say, we can tune the magic wavelengths here by the external parameters of light. Since the wavelength of the  resonance transitions $5s_{\frac{1}{2}} \rightarrow 4d_{\frac{3}{2}}, 4d_{\frac{5}{2}}$ are 687 nm and 674 nm,  all the  magic wavelengths in the visible and UV regions as shown  in the tables, support blue-detuned trapping scheme. Therefore these wavelengths seek the ion to confine in the  low intensity region of the LG beam \cite{Kennedy2014, Kuga1997, Wright2000, Arlt2001} (see FIG. 1). However,   there are few cases where optical magic wavelengths   are larger than these resonance transition wavelengths and support red-detuning trapping. They are 696.69 nm, 694.56nm, 688.27nm (for OAM=$+1$  as shown in TABLE V, and  696.69 nm, 692.45 nm, 682.09 nm (for OAM=$+2$) as shown in TABLE VI.  The same situation appears in case of SAM=$-1$ but when the magnetic component of the final state is $m_J=+5/2$.

As our main focus in this present work is on the magic wavelengths, therefore, we give a rough estimation of the theoretical uncertainty in the calculated magic wavelength values only. Here we recalculate the magic wavelengths (of all in Table III, IV, V and VI) by replacing our present RCC values of the most important $E1$ matrix elements by the corresponding SDpT values as calculated by Safronova \cite{Safronova2010}. These most important $E1$ matrix elements include $5s_{\frac{1}{2}} \rightarrow 5p_{\frac{1}{2}, \frac{3}{2}}$ transitions for $5s_{\frac{1}{2}}$ state; $4d_{\frac{3}{2}} \rightarrow 5p_{\frac{1}{2}, \frac{3}{2}}$ and $4d_{\frac{3}{2}} \rightarrow 4f_{\frac{5}{2}}$  transitions for $4d_{\frac{3}{2}}$ state; $4d_{\frac{5}{2}} \rightarrow 5p_{\frac{3}{2}}$ and $4d_{\frac{5}{2}} \rightarrow 4f_{\frac{5}{2}, \frac{7}{2}}$ transitions for $4d_{\frac{5}{2}}$ state. The maximum of the relative differences between these recalculated wavelengths and the corresponding actual wavelengths (as presented in Table III, IV,  V, and VI) is $\pm$1\% and is considered as the theoretical uncertainty in our calculated magic wavelength values.

\section{CONCLUSIONS}

A theoretical formalism of the dynamic polarizability of an atomic state due to  the LG beam has been presented here in the robust form of the external field keeping in mind  the trapping process of atoms or ions as best possible application. The sum-over-states technique is used to estimate the polarizability values. The  correlation  exhaustive RCC theory is applied to calculate the most important and correlation sensitive  dipole matrix elements inside the sum. List of recommended magic wavelengths in the wide range electromagnetic spectrum, from IR to UV range, is presented for  $5s_{\frac{1}{2}}$ to $4d_{\frac{3}{2}, \frac{5}{2}}$ transition of the Sr$^+$ ion. These will help  to trap the atom or ion in high precision experiments using  both red- and blue-detuned  techniques. Appreciable variations of magic wavelengths with the OAM, SAM and focusing angles of the LG beam are evaluated, and they add extra freedom in the high precision confinement approach as tunability of trapping field.  For comparison, we also present our calculated  magic wavelengths of the above particular transitions using circularly   and  linearly polarized Gaussian beams. The new near-infrared magic wavelengths with high polarizabilities are prescribed as the best wavelengths for trapping. 

\section{ACKNOWLEDGMENTS}
A.B. is thankful to Subrata Das from Indian Institute of Technology Kharagpur for the schematic representation.

\clearpage

\begin{figure*}[!h]
\subfloat[]{\includegraphics[trim = 1cm 3.0cm 0.1cm 3.5cm, scale=.40]{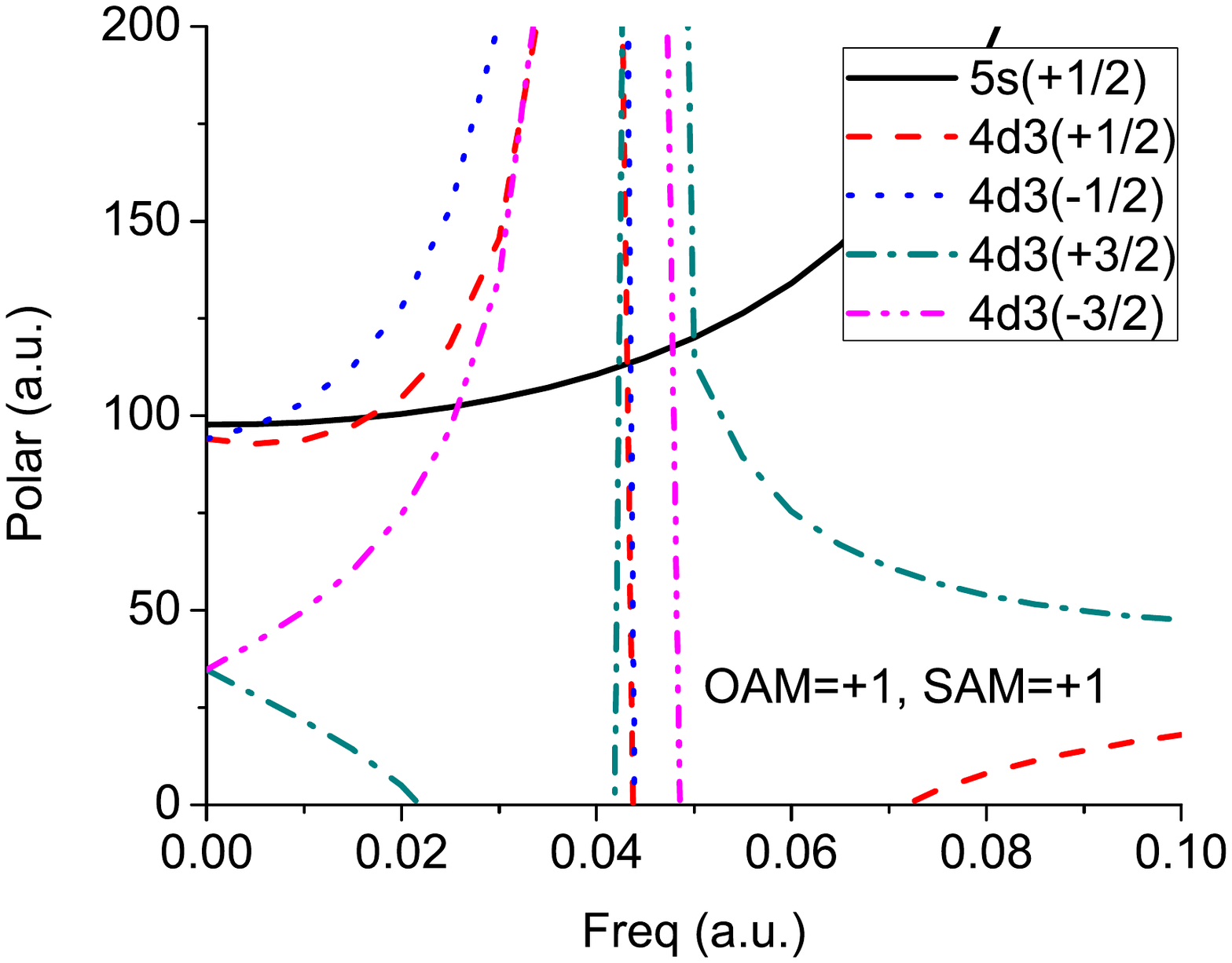}}
\subfloat[]{\includegraphics[trim = 1cm 3.0cm 0.1cm 3.5cm, scale=.40]{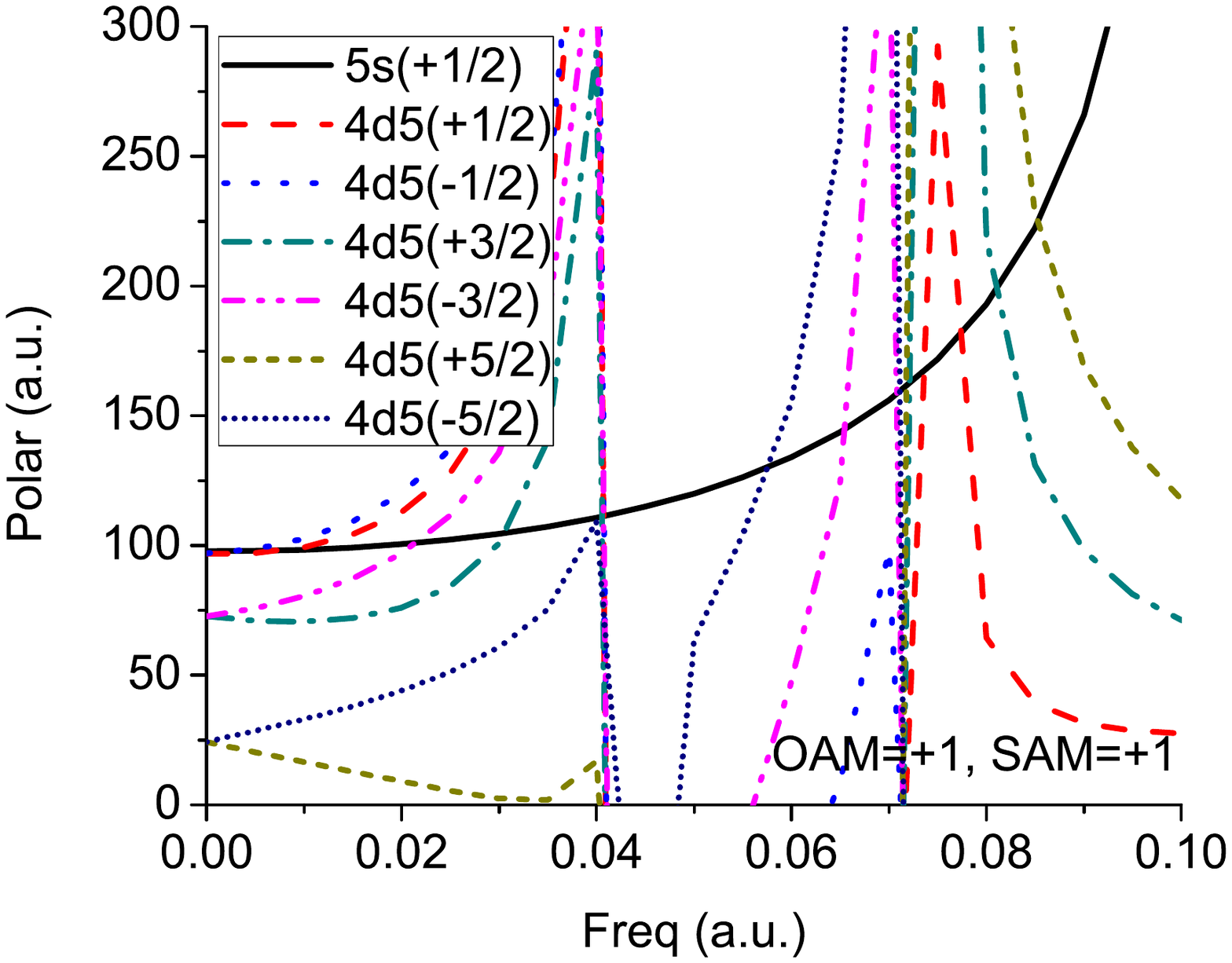}}\\
\subfloat[]{\includegraphics[trim =  1cm 3.0cm 0.1cm 3.1cm,scale=.40]{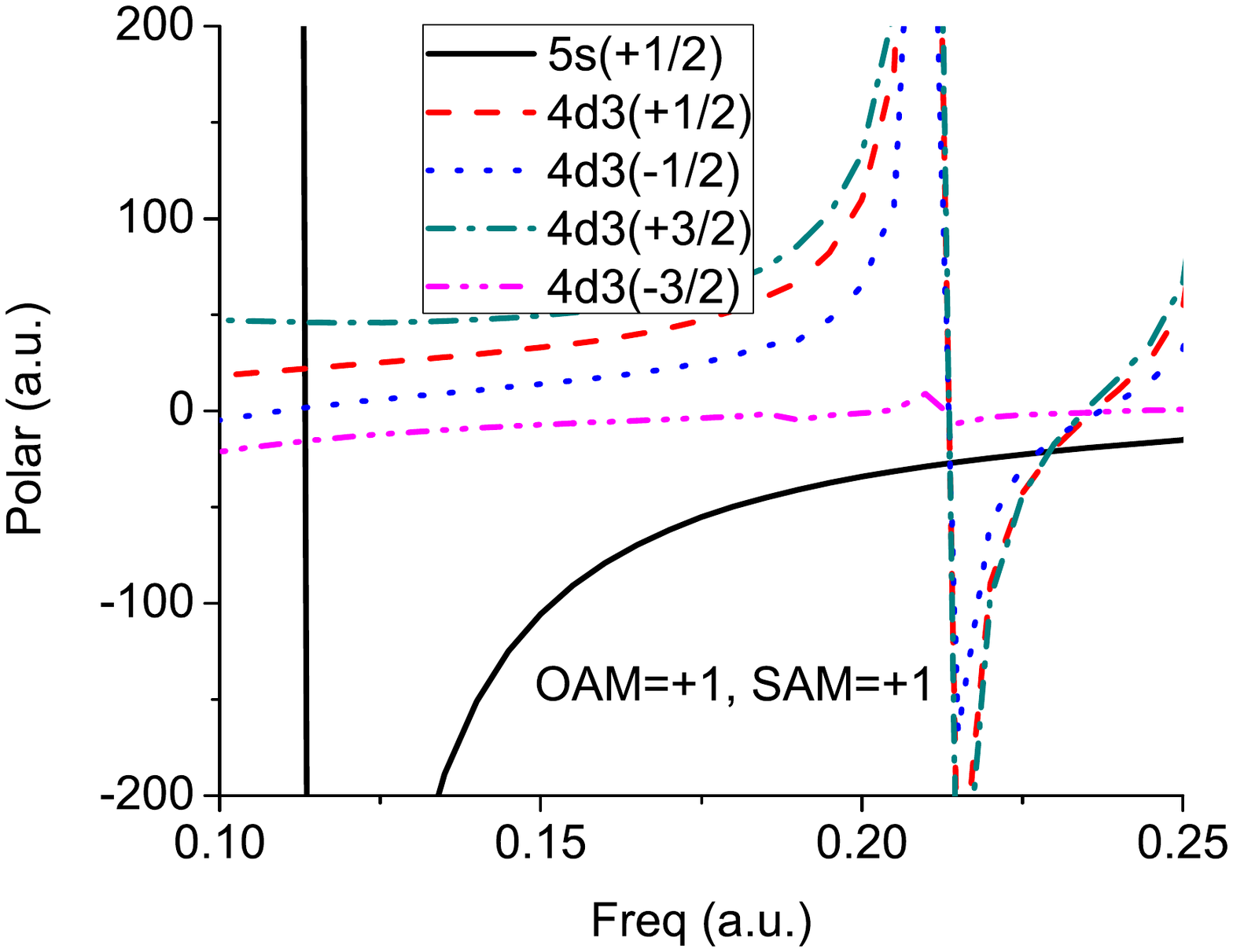}}
\subfloat[]{\includegraphics[trim =  1cm 3.0cm 0.1cm 3.1cm, scale=.40]{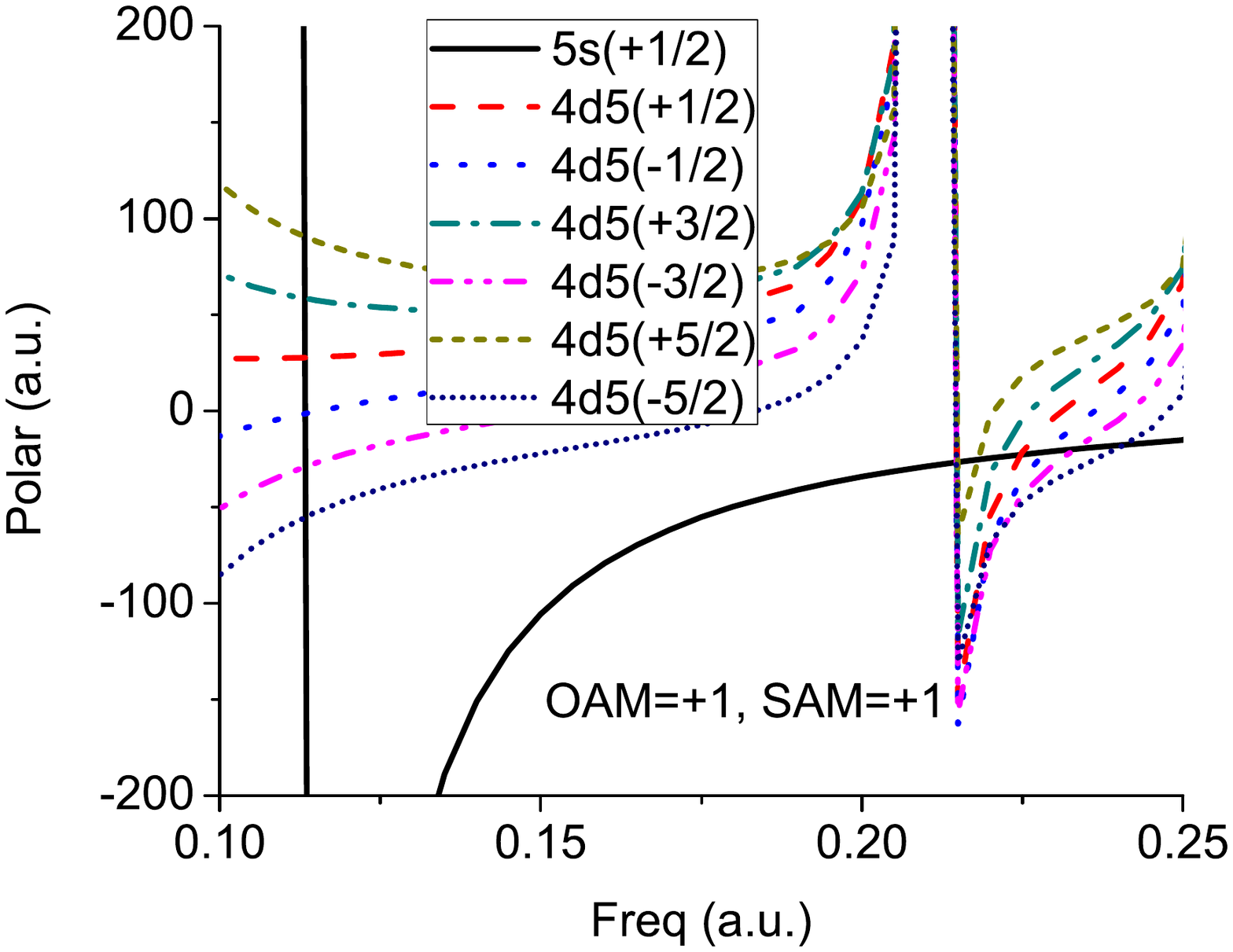}}
\caption{Frequency (Freq) dependence of polarizabilities (Polar) for the $5s_{\frac{1}{2}}$ and $4d_{\frac{3}{2},\frac{5}{2}}$ states for the focused (50$^\circ$) LG beam with OAM=+1 and SAM=+1. The brackets indicate the magnitudes of the magnetic components. Fig. (a) and (c) are for the $5s_{\frac{1}{2}}$ and $4d_{\frac{3}{2}}$ states, and  Fig. (b) and (d) are for the $5s_{\frac{1}{2}}$ and $4d_{\frac{5}{2}}$ states.}
\end{figure*}

\begin{figure*}[!h]
\subfloat[]{\includegraphics[trim = 1cm 3.0cm 0.1cm 3.5cm, scale=.40]{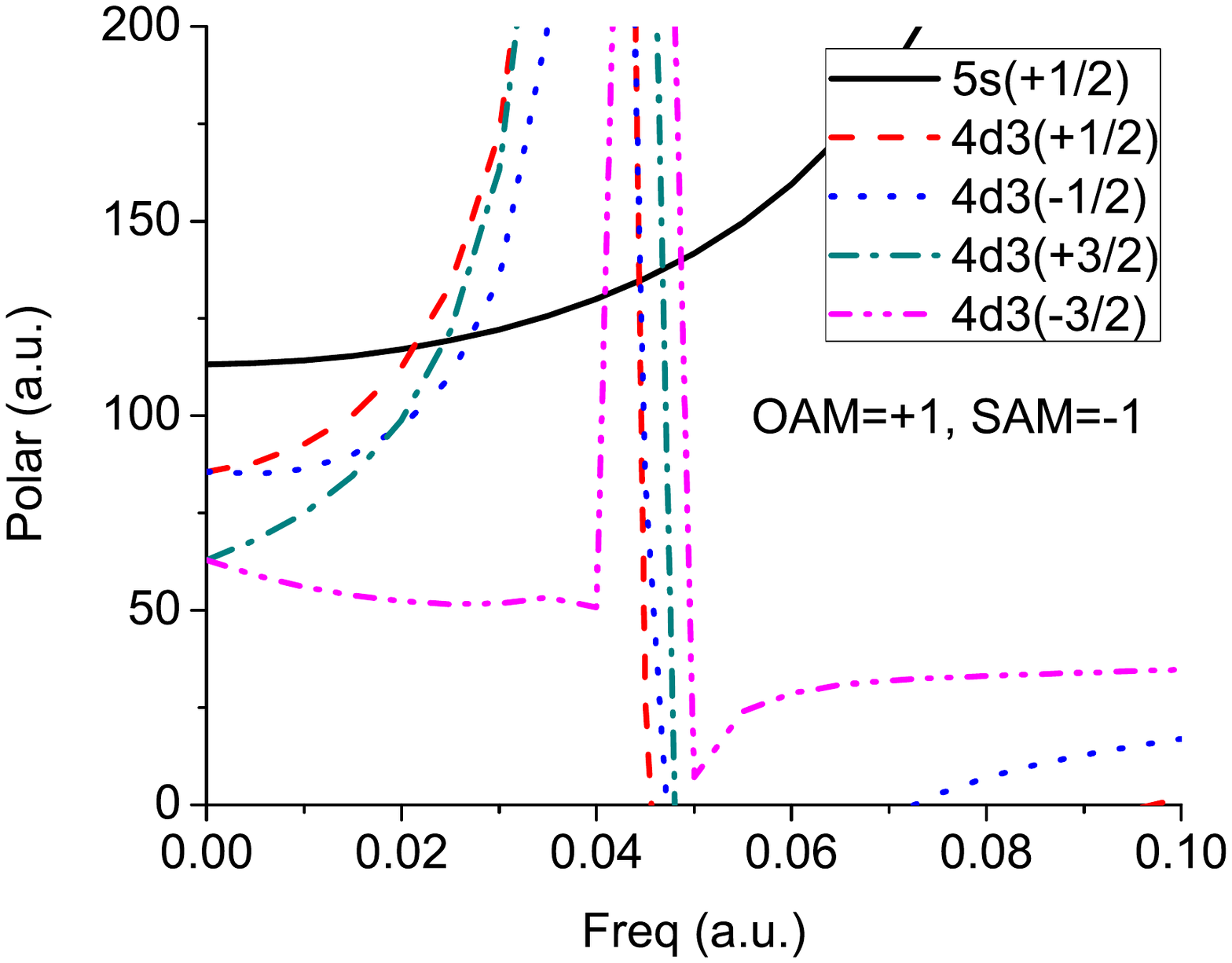}}
\subfloat[]{\includegraphics[trim = 1cm 3.0cm 0.1cm 3.5cm, scale=.40]{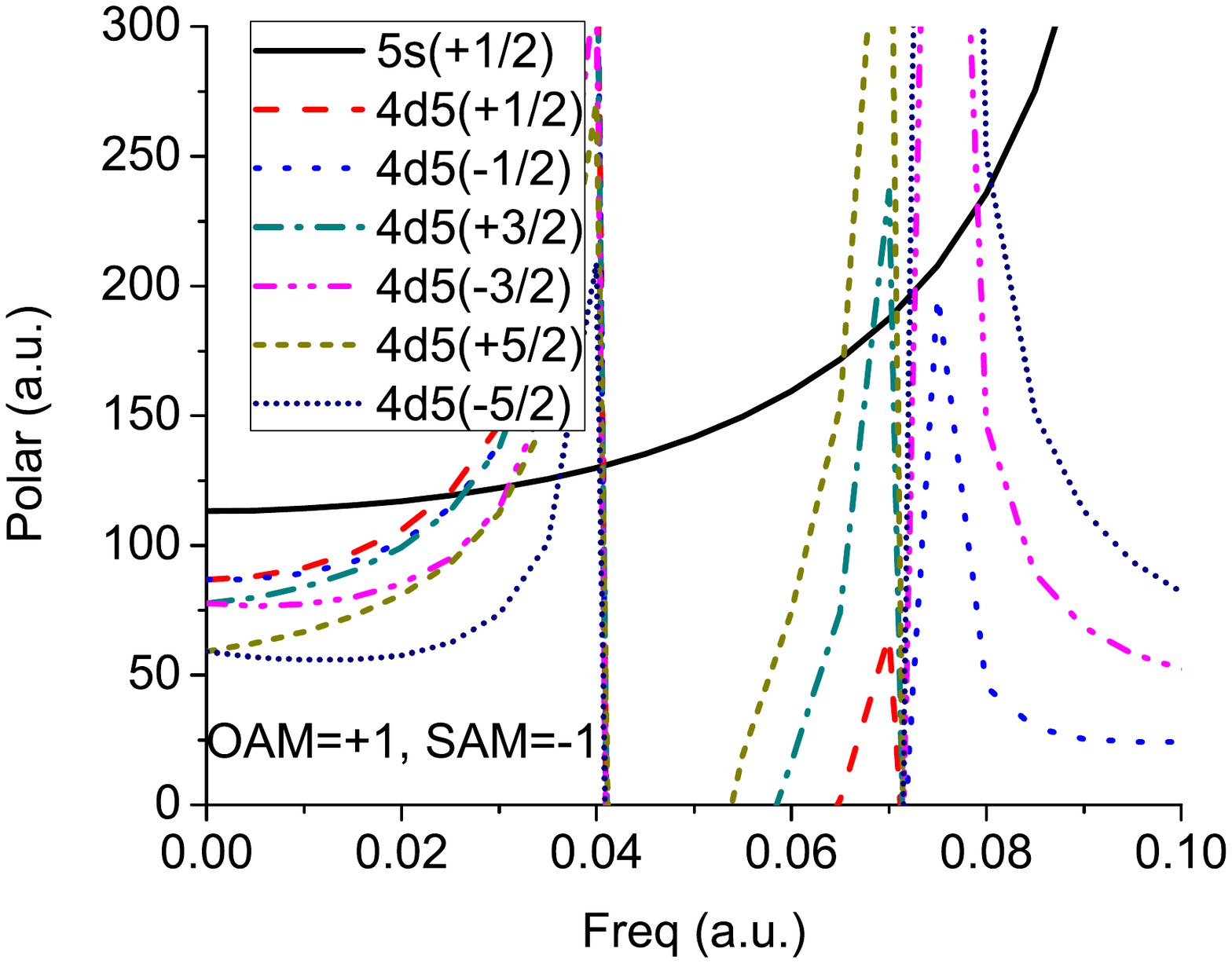}}\\
\subfloat[]{\includegraphics[trim =  1cm 3.0cm 0.1cm 3.1cm,scale=.40]{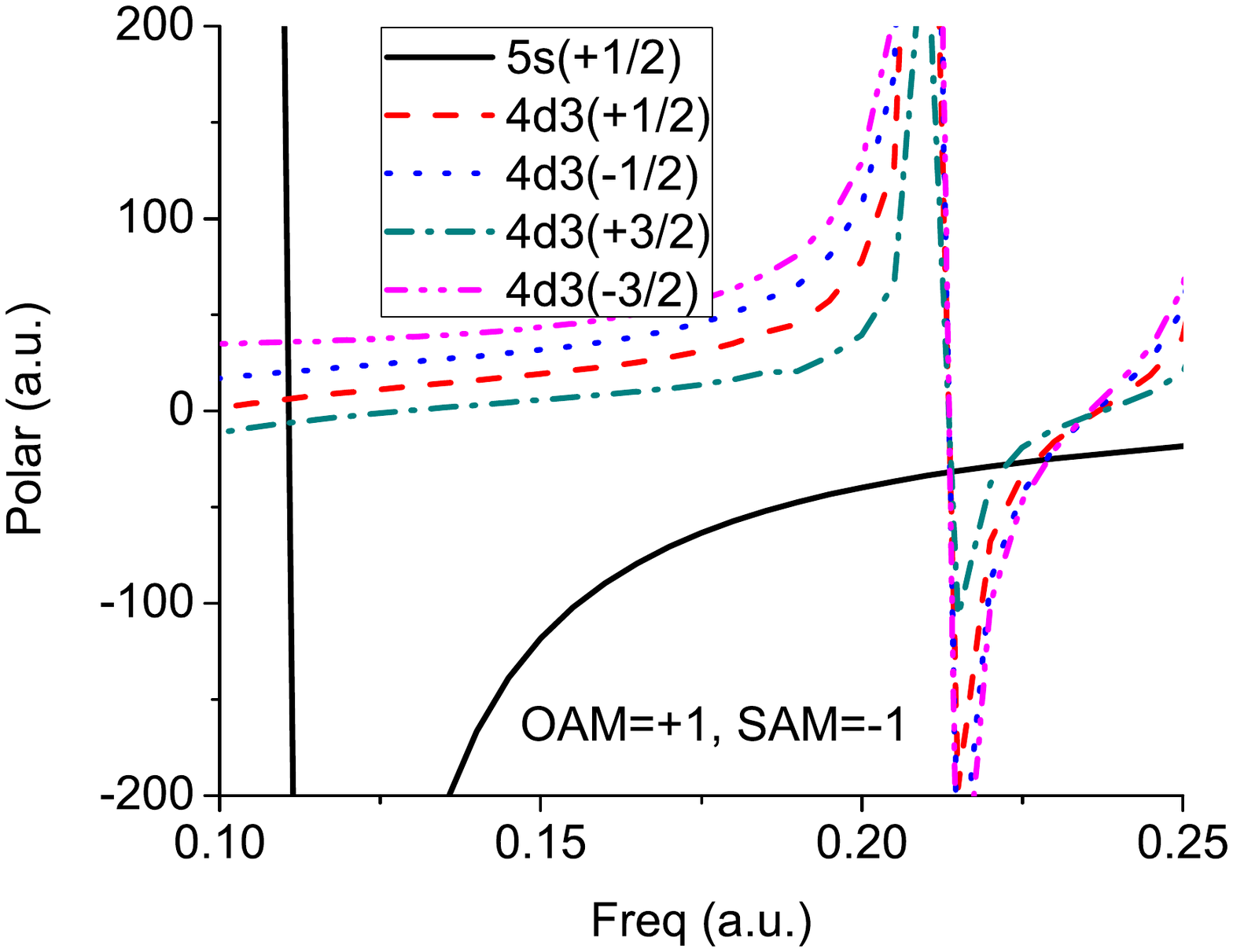}}
\subfloat[]{\includegraphics[trim =  1cm 3.0cm 0.1cm 3.1cm, scale=.40]{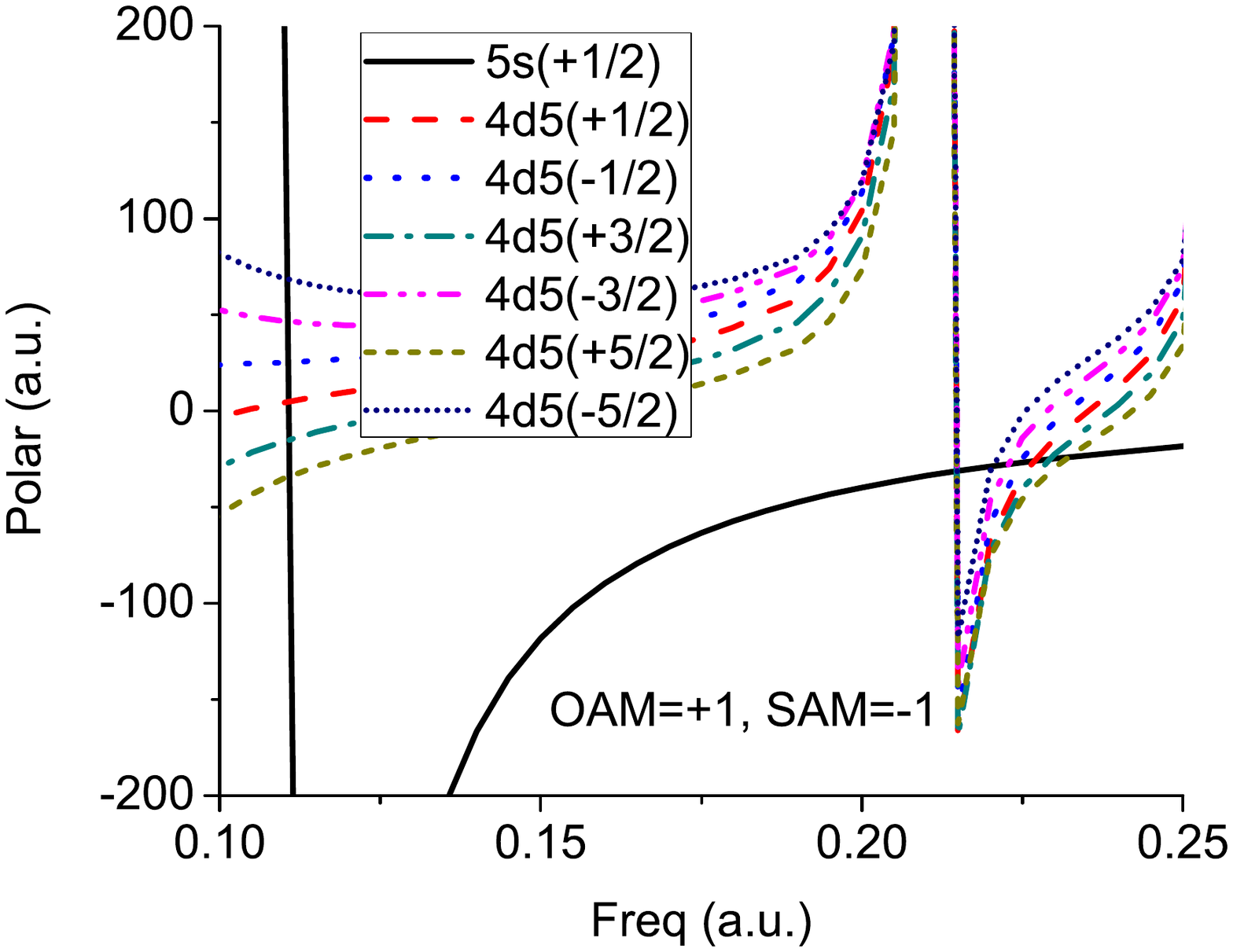}}
\caption{Frequency (Freq) dependence of polarizabilities (Polar) for the $5s_{\frac{1}{2}}$ and $4d_{\frac{3}{2},\frac{5}{2}}$ states for the focused (50$^\circ$) LG beam with OAM=+1 and SAM=$-1$. The brackets indicate the magnitudes of the magnetic components. Fig. (a) and (c) are for the $5s_{\frac{1}{2}}$ and $4d_{\frac{3}{2}}$ states, and  Fig. (b) and (d) are for the $5s_{\frac{1}{2}}$ and $4d_{\frac{5}{2}}$ states.}
\end{figure*}

\begin{table}[h]
\scriptsize
  \caption{Magic wavelengths (in nm)  of Sr$^+$ for different focusing angles of the LG beam for the transitions $5s_{1/2}(+1/2)  $  $\rightarrow$ $4d_{3/2}(m_J)  $.}
\centering
\begin{tabular}{cccccccccccccc}

\hline \hline
     \multicolumn{14}{c}{\textbf{Non-paraxial LG beam}}\\ 

   State ($4d_{3/2}(m_J)$)     &$\lambda_{\textrm{magic}}^{50^\circ}$&$\alpha$ & $\lambda_{\textrm{magic}}^{60^\circ}$&$\alpha$&  $\lambda_{\textrm{magic}}^{70^\circ}$&$\alpha$& State     ($4d_{3/2}(m_J)$) & $\lambda_{\textrm{magic}}^{50^\circ}$&$\alpha$ & $\lambda_{\textrm{magic}}^{60^\circ}$&$\alpha$&  $\lambda_{\textrm{magic}}^{70^\circ}$&$\alpha$
           \\ [0.2ex]
   \hline 
 & \multicolumn{6}{c}{\textbf{OAM=+1, SAM=+1}} && \multicolumn{6}{c}{\textbf{OAM=+1, SAM=-1}}\\  
  \hline  
$(+1/2)  $&	2744.78	&	99.89	&	2255.61	&	104.12	&	2016.08	&	109.02	&	(+1/2)	&	2149.21	&	117.64	&	2090.06	&	118.56	&	1998.39	&	120.88	\\
&	1057.15	&	113.06	&	1045.03	&	118.00	&	1040.26	&	121.99	&		&	1028.52	&	134.88	&	1019.31	&	136.08	&	1023.90	&	137.63	\\

&	403.57	&	21.93	&	404.65	&	21.49	&	405.01	&	21.58	&	
&	409.74	&	6.54	&	408.27	&	7.18	&	406.09	&	8.17	\\
&	213.11	&	-26.69	&	213.01	&	33.33	&	213.21	&	-29.70	&	&	213.11	&	-31.77	&	212.91	&	-32.02	&	213.21	&	-32.84	\\
&	198.45	&	-21.03	&	198.62	&	-21.99	&	198.88	&	-23.18	&	&	200.63	&	-25.95	&	200.63	&	-26.39	&	200.63	&	-26.70	\\

$(-1/2)  $&	8933.99	&	97.82	&	4952.54	&	101.80	&	3301.69	&	109.02	&	$(-1/2)  $	&	1693.80	&	120.16	&	1656.85	&	122.42	&	1644.89	&	123.88	\\
&	1047.43	&	113.58	&	1042.64	&	117.78	&	1042.64	&	122.42	&		&	1023.90	&	134.88	&	1019.31	&	136.08	&	1014.77	&	137.63	\\

&	403.57	&	1.24	&	404.65	&	1.94	&	404.65	&	3.47	&	&	410.11	&	20.48	&	408.27	&	20.30	&	406.45	&	20.19	\\
&	213.11	&	-26.69	&	213.01	&	33.33	&	213.21	&	-29.70	&		&	213.11	&	-31.77	&	212.91	&	-32.02	&	213.21	&	-32.84	\\
&	200.37	&	-21.80	&	200.54	&	-22.65	&	200.63	&	-23.91	&		&	199.23	&	-25.35	&	199.31	&	-25.66	&	199.40	&	-26.23	\\

$(+3/2)  $&	1084.84	&	112.54	&	1077.15	&	116.32	&	1087.43	&	120.10	&	$(+3/2)  $	&	1859.73	&	119.13	&	1859.73	&	120.88	&	1844.67	&	122.42	\\
&	911.27	&	120.16	&	914.93	&	123.88	&	920.47	&	128.09	&		&	973.58	&	137.92	&	977.75	&	139.52	&	977.75	&	140.64	\\

 &	403.93	&	45.95	&	405.01	&	44.78	&	405.01	&	41.99	&		&	409.37	&	-6.02	&	407.91	&	-4.47	&	406.09	&	-2.76	\\
&	213.11	&	-26.69	&	213.01	&	33.33	&	213.21	&	-29.70	&		&	213.11	&	-31.77	&	212.91	&	-32.02	&	213.21	&	-32.84	\\
&	198.79	&	-21.15	&	198.79	&	-22.12	&	198.79	&	-22.92	&		&	204.60	&	-27.80	&	204.14	&	-27.88	&	203.68	&	-28.24	\\

$(-3/2)  $&	1772.89	&	102.40	&	1786.80	&	105.67	&	1815.27	&	109.45	&	$(-3/2)  $	&	1111.30	&	130.81	&	1133.42	&	131.53	&	1195.89	&	131.53	\\
&	953.21	&	117.64	&	957.21	&	120.88	&	961.25	&	125.43	&		&	937.52	&	139.46	&	937.52	&	141.41	&	941.39	&	142.96	\\

&	403.57	&	-16.16	&	404.29	&	-14.23	&	404.65	&	-11.65	&		&	410.48	&	35.78	&	408.64	&	34.49	&	406.82	&	33.09	\\
&	-	&	-	&	-	&	-	&	212.61	&	-28.10	&		&	213.11	&	-31.77	&	212.91	&	-32.02	&	213.21	&	-32.84	\\
&	-	&	-	&	-	&	-	&	208.62	&	-26.51	&		&	198.97	&	-25.21	&	198.97	&	-25.39	&	199.05	&	-26.07	\\

 \hline 
 & \multicolumn{6}{c}{\textbf{OAM=+2, SAM=+1}} && \multicolumn{6}{c}{\textbf{OAM=+2, SAM=-1}}\\  
  \hline

$(+1/2)  $&	2531.30	&	100.86	&	2129.13	&	106.79	&	1815.27	&	113.57	&	$(+1/2)  $	&	2129.13	&	118.61	&	2061.69	&	120.10	&	1963.94	&	121.31	\\
&	1057.15	&	115.06	&	1052.27	&	120.10	&	1042.64	&	126.20	&		&	1023.90	&	135.91	&	1023.90	&	136.86	&	1014.77	&	139.18	\\

&	405.01	&	21.92	&	404.65	&	21.68	&	405.01	&	21.35	&		&	409.37	&	6.63	&	407.18	&	7.66	&	406.45	&	8.58	\\
&	213.31	&	-27.49	&	213.11	&	-28.01	&	213.21	&	-30.27	&		&	213.21	&	-31.35	&	213.01	&	-32.78	&	213.21	&	-33.25	\\
&	198.45	&	-21.44	&	198.71	&	-22.62	&	198.97	&	-24.35	&		&	200.63	&	-26.10	&	200.72	&	-26.58	&	200.72	&	-26.94	\\

$(-1/2)  $&	6603.38	&	99.37	&	3828.85	&	103.35	&	2761.42	&	109.45	&	$(-1/2)  $	&	1662.90	&	121.19	&	1656.85	&	122.42	&	1633.10	&	125.00	\\
&	1047.43	&	115.58	&	1045.03	&	120.10	&	1042.64	&	126.20	&		&	1023.90	&	135.62	&	1023.90	&	136.86	&	1010.27	&	139.52	\\

&	404.65	&	1.40	&	404.65	&	3.07	&	404.65	&	4.77	&		&	409.74	&	20.45	&	407.54	&	20.34	&	406.45	&	19.94	\\
&	213.31	&	-27.49	&	213.11	&	-28.01	&	213.21	&	-30.27	&		&	213.21	&	-31.35	&	213.01	&	-32.78	&	213.21	&	-33.25	\\
&	200.45	&	-22.04	&	200.54	&	-23.37	&	200.72	&	-24.64	&		&	199.31	&	-25.54	&	199.40	&	-25.90	&	199.58	&	-26.50	\\

$(+3/2)  $&	1082.26	&	113.57	&	1082.26	&	117.78	&	1092.65	&	124.31	&	$(+3/2)  $	&	1859.73	&	119.87	&	1867.35	&	121.65	&	1844.67	&	122.42	\\
&	914.93	&	121.19	&	918.62	&	126.20	&	926.08	&	131.53	&		&	973.58	&	137.92	&	979.86	&	139.52	&	977.75	&	141.41	\\

&	405.01	&	45.54	&	405.01	&	43.71	&	405.01	&	40.32	&		&	409.37	&	-5.20	&	407.18	&	-3.70	&	406.09	&	-1.75	\\
&	213.31	&	-27.49	&	213.11	&	-28.01	&	213.21	&	-30.27	&		&	213.21	&	-31.35	&	213.01	&	-32.78	&	213.21	&	-33.25	\\
&	198.79	&	-21.44	&	198.79	&	-22.62	&	198.97	&	-24.35	&		&	204.50	&	-27.86	&	203.95	&	-28.01	&	203.41	&	-28.29	\\

$(-3/2)  $&	1766.02	&	103.89	&	1808.07	&	107.13	&	1837.23	&	112.46	&	$(-3/2)  $	&	1119.49	&	131.56	&	1141.94	&	132.65	&	1238.13	&	130.76	\\
&	953.21	&	119.13	&	957.21	&	123.88	&	965.33	&	129.64	&		&	935.59	&	139.98	&	945.30	&	141.41	&	947.26	&	142.96	\\

&	404.65	&	-15.26	&	404.65	&	-13.11	&	404.65	&	-9.99	&		&	410.11	&	35.50	&	407.91	&	33.94	&	406.82	&	32.17	\\
&	-	&	-	&	212.22	&	-28.01	&	212.71	&	-29.97	&		&	213.21	&	-31.35	&	213.01	&	-32.78	&	213.21	&	-33.25	\\
&	-	&	-	&	209.97	&	-28.01	&	207.58	&	-27.62	&		&	198.97	&	-25.36	&	199.05	&	-25.89	&	199.14	&	-26.31	\\

\\

\hline
\hline
\label{table:nonlin} 
\label{I}

\end{tabular}
\end{table}

\begin{table}[h]
\scriptsize
  \caption{Magic wavelengths (in nm)  of Sr$^+$ for different focusing angles of the LG beam for the transitions $5s_{1/2}(+1/2)  $  $\rightarrow$ $4d_{5/2}(m_J)$.}
\centering
\begin{tabular}{cccccccccccccc}

\hline \hline
     \multicolumn{14}{c}{\textbf{Non-paraxial LG beam}}\\ 

   State ($4d_{5/2}(m_J)$)     & $\lambda_{\textrm{magic}}^{50^\circ}$&$\alpha$ & $\lambda_{\textrm{magic}}^{60^\circ}$&$\alpha$&  $\lambda_{\textrm{magic}}^{70^\circ}$&$\alpha$& State     ($4d_{5/2}(m_J)$) & $\lambda_{\textrm{magic}}^{50^\circ}$&$\alpha$ & $\lambda_{\textrm{magic}}^{60^\circ}$&$\alpha$&  $\lambda_{\textrm{magic}}^{70^\circ}$&$\alpha$
           \\ [0.2ex]
   \hline 
 & \multicolumn{6}{c}{\textbf{OAM=+1, SAM=+1}} && \multicolumn{6}{c}{\textbf{OAM=+1, SAM=-1}}\\  
  \hline  
$(+1/2)  $&	7855.75	&	98.02	&	3057.94	&	101.80	&	2348.63	&	107.13	&	$(+1/2)  $	&	1844.67	&	118.56	&	1815.27	&	120.10	&	1766.02	&	122.42	\\
&	1119.49	&	111.77	&	1119.49	&	114.78	&	1119.80	&	119.33	&		&	1116.75	&	130.76	&	1111.30	&	133.08	&	1119.49	&	133.08	\\
&	619.91	&	166.58	&	618.23	&	173.37	&	614.89	&	181.01	&		&	-	&	-	&	-	&	-	&	-	&	-	\\
&	588.67	&	182.13	&	590.20	&	186.34	&	594.82	&	191.67	&		&	-	&	-	&	-	&	-	&	-	&	-	\\

&	403.57	&	26.86	&	403.57	&	27.57	&	403.57	&	27.36	&		&	411.59	&	4.61	&	410.85	&	5.83	&	410.11	&	6.70	\\
&	212.32	&	-27.12	&	212.12	&	-27.54	&	212.12	&	-29.02	&		&	211.92	&	-31.00	&	212.12	&	-31.65	&	212.12	&	-32.35	\\
&	202.68	&	-22.49	&	202.77	&	-23.84	&	202.77	&	-24.62	&		&	200.81	&	-26.03	&	200.90	&	-26.49	&	201.07	&	-26.89	\\

$(-1/2)  $&	8136.31	&	98.02	&	4602.36	&	102.23	&	2830.02	&	106.44	&	$(-1/2)  $	&	1732.45	&	120.87	&	1719.37	&	121.65	&	1681.30	&	123.20	\\
&	1119.49	&	111.00	&	1114.02	&	114.43	&	1119.49	&	119.33	&		&	1116.75	&	130.76	&	1111.30	&	133.08	&	1119.49	&	133.08	\\

&	403.22	&	-1.31	&	403.22	&	-0.14	&	403.57	&	2.33	&		&	411.97	&	25.26	&	410.85	&	24.66	&	410.11	&	23.90	\\
&	212.32	&	-27.12	&	212.12	&	-27.54	&	212.12	&	-29.02	&		&	211.92	&	-31.00	&	212.12	&	-31.65	&	212.12	&	-32.35	\\
&	199.23	&	-21.45	&	199.49	&	-22.34	&	199.93	&	-23.35	&		&	202.77	&	-26.88	&	202.86	&	-27.33	&	202.77	&	-26.68	\\

$(+3/2)  $&	1493.88	&	104.90	&	1493.88	&	107.90	&	1493.88	&	111.77	&	$(+3/2)  $	&	1732.45	&	120.87	&	1719.37	&	121.65	&	1693.80	&	123.20	\\
&	1130.60	&	111.00	&	1125.02	&	114.43	&	1125.02	&	119.33	&		&	1116.75	&	130.76	&	1111.30	&	133.08	&	1119.49	&	133.08	\\
&	631.95	&	161.94	&	631.95	&	168.04	&	631.95	&	174.91	&		&	667.11	&	181.79	&	661.30	&	184.79	&	-	&	-	\\
&	561.82	&	200.00	&	568.12	&	201.55	&	570.97	&	206.87	&		&	648.13	&	188.23	&	648.13	&	190.12	&	-	&	-	\\

 &	403.57	&	58.56	&	403.57	&	56.67	&	403.93	&	53.79	&		&	411.22	&	-15.08	&	410.48	&	-12.64	&	409.74	&	-9.97	\\
&	212.32	&	-27.12	&	212.12	&	-27.54	&	212.12	&	-29.02	&		&	211.92	&	-31.00	&	212.12	&	-31.65	&	212.12	&	-32.35	\\
&	205.80	&	-23.96	&	205.52	&	-24.96	&	205.24	&	-25.84	&		&	198.79	&	-25.18	&	199.05	&	-25.64	&	199.31	&	-26.13	\\

$(-3/2)  $&	2149.21	&	101.12	&	2034.08	&	104.90	&	1922.50	&	108.68	&	$(-3/2)  $	&	1474.54	&	123.2	&	1474.54	&	124.66	&	1479.33	&	125.43	\\
&	1119.49	&	111.00	&	1114.02	&	114.78	&	1119.49	&	118.56	&		&	1119.49	&	130.76	&	1111.30	&	133.08	&	1119.49	&	133.08	\\
&	696.69	&	144.50	&	694.56	&	150.52	&	688.27	&	158.16	&		&	625.87	&	197.77	&	625.87	&	200.00	&	625.01	&	203.44	\\
&	646.29	&	157.39	&	645.37	&	163.48	&	644.46	&	169.59	&		&	576.75	&	229.73	&	576.75	&	232.73	&	579.69	&	231.62	\\

&	403.22	&	-29.48	&	403.22	&	-26.38	&	403.22	&	-22.77	&		&	411.97	&	46.39	&	411.22	&	44.70	&	410.48	&	42.60	\\
&	212.32	&	-27.12	&	212.12	&	-27.54	&	212.12	&	-29.02	&		&	211.92	&	-31.00	&	212.12	&	-31.65	&	212.12	&	-32.35	\\
&	195.30	&	-19.78	&	196.14	&	-20.85	&	197.07	&	-22.13	&		&	204.41	&	-27.49	&	204.23	&	-28.00	&	204.14	&	-28.38	\\

$(+5/2)  $&		-&	-	&	-	&	-	&	-	&		-&	$(+5/2)  $	&	1451.06	&	123.2	&	1474.54	&	124.66	&	1479.33	&	125.43	\\
&	-	&	-	&	-	&	-	&	-	&	-	&		&	1119.49	&	130.76	&	1111.30	&	133.08	&	1119.49	&	133.08	\\
&	635.47	&	160.48	&	633.70	&	166.92	&	633.70	&	174.91	&		&	696.69	&	172.59	&	694.56	&	175.69	&	690.35	&	178.69	\\
&	534.15	&	224.40	&	539.21	&	227.41	&	544.37	&	231.96	&		&	644.46	&	189.35	&	643.55	&	192.43	&	644.46	&	193.90	\\

&	403.93	&	91.63	&	403.93	&	87.23	&	403.93	&	82.21	&		&	411.22	&	-34.28	&	410.11	&	-30.69	&	409.74	&	-26.69	\\
&	212.32	&	-27.12	&	212.12	&	-27.54	&	212.12	&	-29.02	&		&	211.92	&	-31.00	&	212.12	&	-31.65	&	212.12	&	-32.35	\\
&	208.91	&	-25.08	&	208.43	&	-26.06	&	207.96	&	-27.11	&		&	196.22	&	-24.13	&	196.82	&	-24.61	&	197.50	&	-25.31	\\

$(-5/2)  $&		-&	-	&	1238.13	&	111.77	&	1279.87	&	114.78	&	$(-5/2)  $	&	1258.66	&	127.32	&	1276.28	&	127.75	&	1287.10	&	129.21	\\
&	-	&		-&	1130.60	&	114.78	&	1125.02	&	119.33	&		&	1125.02	&	130.76	&	1111.30	&	133.08	&	1119.49	&	133.08	\\
&	795.17	&	129.98	&	759.39	&	138.4	&	743.28	&	146.74	&		&	631.95	&	195.45	&	632.82	&	196.99	&	628.46	&	200	\\
&	641.74	&	159.70	&	642.64	&	164.26	&	640.83	&	170.7	&		&	565.30	&	239.6	&	568.12	&	237.29	&	570.97	&	238.83	\\

&	402.86	&	-55.51	&	403.22	&	-51.23	&	403.22	&	-46.48	&		&	412.34	&	68.54	&	411.22	&	65.29	&	410.48	&	61.25	\\
&	212.32	&	-27.12	&	212.12	&	-27.54	&	212.12	&	-29.02	&		&	211.92	&	-31.00	&	212.12	&	-31.65	&	212.12	&	-32.35	\\
&	189.22	&	-17.49	&	190.72	&	-19	&	192.74	&	-20.86	&		&	206.73	&	-28.55	&	206.26	&	-28.84	&	205.89	&	-29.21	\\

\hline
\hline
\label{table:nonlin} 
\label{I}

\end{tabular}
\end{table}

\begin{table}[h]
\scriptsize
  \caption{Magic wavelengths (in nm)  of Sr$^+$ for different focusing angles of the LG beam for the transitions $5s_{1/2}(+1/2)  $  $\rightarrow$ $4d_{5/2}(m_J)  $.}
\centering
\begin{tabular}{cccccccccccccc}

\hline \hline
     \multicolumn{14}{c}{\textbf{Non-paraxial LG beam}}\\ 

   State  ($4d_{5/2}(m_J)$)    & $\lambda_{\textrm{magic}}^{50^\circ}$&$\alpha$ & $\lambda_{\textrm{magic}}^{60^\circ}$&$\alpha$&  $\lambda_{\textrm{magic}}^{70^\circ}$&$\alpha$& State     ($4d_{5/2}(m_J)$) & $\lambda_{\textrm{magic}}^{50^\circ}$&$\alpha$ & $\lambda_{\textrm{magic}}^{60^\circ}$&$\alpha$&  $\lambda_{\textrm{magic}}^{70^\circ}$&$\alpha$
           \\ [0.2ex]

 \hline 
 & \multicolumn{6}{c}{\textbf{OAM=+2, SAM=+1}} && \multicolumn{6}{c}{\textbf{OAM=+2, SAM=-1}}\\  
  \hline

$(+1/2)  $&	4423.63	&	98.8	&	2618.58	&	104.90	&	2090.06	&	111.00	&	$(+1/2)  $	&	1859.73	&	119.76	&	1786.80	&	120.88	&	1732.45	&	123.20	\\
&	1116.75	&	113.23	&	1119.49	&	117.10	&	1125.02	&	123.54	&		&	1119.49	&	130.76	&	1125.02	&	133.08	&	1119.49	&	134.54	\\
&	619.91	&	169.59	&	616.55	&	177.92	&	614.89	&	188.23	&		&	-	&	-	&	-	&	-	&	-	&	-	\\
&	588.67	&	184.02	&	590.96	&	189.35	&	597.94	&	196.99	&		&	-	&	-	&	-	&	-	&	-	&	-	\\

&	403.93	&	27.47	&	404.29	&	27.62	&	404.65	&	26.83	&		&	411.22	&	4.98	&	410.11	&	6.35	&	408.64	&	7.99	\\
&	212.02	&	-26.57	&	212.12	&	-28.04	&	212.02	&	-30.53	&		&	212.42	&	-31.78	&	212.22	&	-32.35	&	211.92	&	-32.89	\\
&	202.68	&	-23.14	&	202.68	&	-24.09	&	202.86	&	-25.69	&		&	200.81	&	-26.16	&	200.99	&	-26.68	&	201.25	&	-27.25	\\

$(-1/2)  $&	9492.37	&	98.02	&	3325.79	&	104.12	&	2373.09	&	110.22	&	$(-1/2)  $		&	1732.45	&	121.22	&	1693.80	&	122.42	&	1675.12	&	123.88	\\
&	1116.75	&	113.23	&	1119.49	&	117.10	&	1125.02	&	123.54	&		&	1119.49	&	130.76	&	1125.02	&	133.08	&	1119.49	&	134.54	\\

&	403.57	&	-0.98	&	404.29	&	0.77	&	404.65	&	3.17	&	&	411.59	&	25.45	&	410.48	&	24.64	&	409.01	&	23.95	\\
&	212.02	&	-26.57	&	212.12	&	-28.04	&	212.02	&	-30.53	&		&	212.42	&	-31.78	&	212.22	&	-32.35	&	211.92	&	-32.89	\\
&	199.40	&	-21.73	&	199.75	&	-23.09	&	200.28	&	-24.60	&		&	202.86	&	-27.33	&	202.86	&	-27.61	&	202.86	&	-27.79	\\

$(+3/2)  $&	1493.88	&	106.44	&	1493.88	&	110.22	&	1479.33	&	115.21	&	$(+3/2)  $	&	1719.37	&	121.22	&	1693.80	&	122.42	&	1675.12	&	123.88	\\
&	1116.75	&	113.23	&	1119.49	&	117.10	&	1125.02	&	123.54	&		&	1119.49	&	130.76	&	1125.02	&	133.08	&	1119.49	&	134.54	\\
&	633.70	&	164.69	&	631.95	&	171.91	&	631.95	&	180.67	&		&	667.11	&	181.79	&	659.38	&	186.68	&	-	&	-	\\
&	565.30	&	200.77	&	570.97	&	205.33	&	572.40	&	211.43	&		&	646.29	&	190.89	&	649.98	&	191.24	&	-	&	-	\\

&	403.93	&	58.58	&	404.29	&	55.64	&	404.65	&	51.65	&		&	411.22	&	-14.05	&	410.11	&	-11.07	&	408.64	&	-7.91	\\
&	212.02	&	-26.57	&	212.12	&	-28.04	&	212.02	&	-30.53	&		&	212.42	&	-31.78	&	212.22	&	-32.35	&	211.92	&	-32.89	\\
&	205.70	&	-24.37	&	205.33	&	-25.48	&	205.06	&	-26.51	&		&	198.88	&	-25.41	&	199.23	&	-25.75	&	199.58	&	-26.41	\\

$(-3/2)  $&	2090.06	&	101.8	&	1963.94	&	107.13	&	1844.67	&	112.46	&	$(-3/2)  $	&	1474.54	&	123.2	&	1474.54	&	125.09	&	1479.33	&	125.86	\\
&	1116.75	&	113.23	&	1119.49	&	117.10	&	1125.02	&	123.54	&		&	1119.49	&	130.76	&	1125.02	&	133.08	&	1119.49	&	134.54	\\
&	696.69	&	147.51	&	692.45	&	154.73	&	682.09	&	163.49	&		&	626.73	&	198.54	&	626.73	&	201.55	&	626.73	&	204.64	\\
&	642.64	&	160.48	&	644.46	&	167.27	&	647.21	&	175.26	&		&	575.29	&	231.27	&	579.69	&	232.73	&	579.69	&	233.51	\\

&	403.57	&	-28.85	&	403.93	&	-24.91	&	404.29	&	-20.66	&		&	411.59	&	45.92	&	410.48	&	43.73	&	409.01	&	40.98	\\
&	212.02	&	-26.57	&	212.12	&	-28.04	&	212.02	&	-30.53	&		&	212.42	&	-31.78	&	212.22	&	-32.35	&	211.92	&	-32.89	\\
&	195.72	&	-20.32	&	196.48	&	-21.86	&	197.76	&	-23.53	&		&	204.41	&	-27.71	&	204.23	&	-28.06	&	204.05	&	-28.63	\\

$(+5/2)  $&	-	&	-	&	-	&	-	&	1171.29	&	120.10	&$(+5/2)  $		&	1451.06	&	123.2	&	1474.54	&	125.09	&	1479.33	&	125.86	\\
&	-	&	-	&	-	&	-	&	1144.81	&	121.99	&		&	1119.49	&	130.76	&	1125.02	&	133.08	&	1119.49	&	134.54	\\
&	633.70	&	162.71	&	631.95	&	171.91	&	633.70	&	180.24	&		&	696.69	&	174.14	&	692.45	&	177.23	&	686.20	&	181.36	\\
&	536.67	&	225.17	&	541.78	&	230.50	&	548.30	&	233.51	&		&	642.64	&	191.24	&	646.29	&	193.21	&	646.29	&	194.67	\\

&	404.29	&	90.32	&	404.29	&	84.88	&	404.65	&	78.14	&		&	410.85	&	-32.86	&	409.74	&	-28.91	&	408.27	&	-23.69	\\
&	212.02	&	-26.57	&	212.12	&	-28.04	&	212.02	&	-30.53	&		&	212.42	&	-31.78	&	212.22	&	-32.35	&	211.92	&	-32.89	\\
&	208.72	&	-25.58	&	208.15	&	-26.59	&	207.58	&	-27.87	&		&	196.39	&	-24.24	&	197.07	&	-25.06	&	198.02	&	-25.60	\\

$(-5/2)  $&	1180.40	&	110.65	&	1269.17	&	114	&	1340.10	&	118.21	&	$(-5/2)  $	&	1262.14	&	126.98	&	1283.47	&	128.52	&	1287.10	&	129.98	\\
&	1130.60	&	112.45	&	1119.49	&	117.10	&	1125.02	&	123.54	&		&	1119.49	&	130.76	&	1125.02	&	133.08	&	1119.49	&	134.54	\\
&	781.53	&	132.3	&	753.11	&	143.3	&	729.01	&	153.61	&		&	631.95	&	195.45	&	632.82	&	199.23	&	631.07	&	201.55	\\
&	640.83	&	161.94	&	640.83	&	168.81	&	642.64	&	176.03	&		&	568.12	&	239.6	&	569.54	&	238.4	&	572.40	&	240.38	\\

&	403.57	&	-54.76	&	403.93	&	-49.37	&	404.29	&	-42.34	&		&	411.97	&	67.26	&	410.48	&	63.32	&	409.37	&	58.69	\\
&	212.02	&	-26.57	&	212.12	&	-28.04	&	212.02	&	-30.53	&		&	212.42	&	-31.78	&	212.22	&	-32.35	&	211.92	&	-32.89	\\
&	189.61	&	-18.1	&	191.68	&	-19.65	&	194.05	&	-21.88	&		&	206.54	&	-28.87	&	206.17	&	-29.07	&	205.70	&	-29.17	\\

\hline
\hline
\label{table:nonlin} 
\label{I}

\end{tabular}
\end{table}

\clearpage



\end{document}